\documentclass[aps,twocolumn,prc,superscriptaddress,showpacs,nofootinbib,floatfix,amssymb,amsfonts,amsmath]{revtex4-1}

\usepackage{color}
\usepackage{longtable}
\usepackage{graphicx}
\usepackage{dcolumn}
\usepackage{bm}
\usepackage{enumerate}
\usepackage{setspace}
\usepackage[dvipdfm,bookmarks=true,colorlinks,%
            citecolor=blue,linkcolor=blue,hypertex,%
            breaklinks=true]{hyperref}

\begin{document}


\title{Rotational excitations in rare-earth nuclei:
       a comparative study within three cranking models with
       different mean fields and the treatments of pairing correlations}

\author{Zhen-Hua Zhang} 
 \email{zhzhang@ncepu.edu.cn}
 \affiliation{Mathematics and Physics Department,
              North China Electric Power University, Beijing 102206, China}
 \affiliation{Department of Physics and Astronomy, Mississippi
State University, MS 39762, USA}

\author{Miao Huang }
 \affiliation{Mathematics and Physics Department,
              North China Electric Power University, Beijing 102206, China}

\author{A.\ V.\ Afanasjev}
 \email{anatoli.afanasjev@gmail.com}
\affiliation{Department of Physics and Astronomy, Mississippi
             State University, MS 39762, USA}
\affiliation{Yukawa Institute for Theoretical Physics,
             Kyoto University, 606-8502 Kyoto, Japan}

\date{\today}

\begin{abstract}
High-spin rotational bands in rare-earth Er ($Z=68$), Tm ($Z=69$) and Yb ($Z=70$)
isotopes are investigated by three different nuclear models.
These are (i) the cranked relativistic Hartree-Bogoliubov (CRHB) approach with
approximate particle number projection by means of the Lipkin-Nogami (LN) method,
(ii) the cranking covariant density functional theory (CDFT) with pairing correlations
treated by a shell-model-like approach (SLAP) or the so called particle-number conserving
(PNC) method, and (iii) cranked shell model (CSM) based on the Nilsson potential
with pairing correlations treated by the PNC method. A detailed comparison between these
three models in the description of the ground state
rotational bands of even-even Er and Yb isotopes is performed.
The similarities and differences between these models in the description of the moments of inertia,
the features of band crossings, equilibrium deformations and pairing energies of
even-even nuclei under study are discussed.
These quantities are considered as a function of rotational frequency and proton and neutron numbers.
The changes in the properties of the first band crossings with increasing
neutron number in this mass region are investigated.
On average, a comparable accuracy of the description of available experimental data
is achieved in these models.
However, the differences between model predictions become larger above the first band crossings.
Because of time-consuming nature of numerical calculations in the CDFT-based models,
a systematic study of the rotational properties of both ground state
and excited state bands in odd-mass Tm nuclei is carried out only by the PNC-SCM.
With few exceptions, the rotational properties of experimental 1-quasiparticle and
3-quasiparticle bands in $^{165,167,169,171}$Tm are reproduced reasonably well.
The appearance of backbendings or upbendings in these nuclei is well understood from
the analysis of  the variations of the occupation probabilities of the single-particle states
and their contributions to total angular momentum alignment with rotational frequency.
\end{abstract}

\maketitle


\section{Introduction \label{Sec:Intro}}

The increase of angular momentum towards extreme values
triggers the appearance of different physical phenomena such as
backbending~\cite{Johnson1971_PLB34-605, Lee1977_PRL38-1454},
band termination~\cite{Bengtsson1983_PST5-165, Afanasjev1999_PR322-1},
signature inversion~\cite{Bengtsson1984_NPA415-189},
superdeformation~\cite{Twin1986_PRL57-811},
wobbling motion~\cite{Odegard2001_PRL86-5866}, etc.
The rare-earth nuclei with $N\sim100$ and $A\sim 170$ are particularly rich in such phenomena.
In this mass region, the nuclei have prolate shapes at ground states but the
yrast and near-yrast structures at medium and high spin are built by a significant number of
multi-quasiparticle (qp) configurations with different degree of triaxiality.
In even-even nuclei the transition from ground state rotational band to 2-qp band is
triggered by first paired band crossing leading either to backbending or upbending.
The backbending has first been observed in $^{160}$Dy ($Z=66$) in the pioneering work
of Ref.~\cite{Johnson1971_PLB34-605}, and later it was interpreted as the alignment
of one pair of the $i_{13/2}$ neutrons~\cite{Stephens1972_NPA183-257}.
Thus, the determination of the nature of band crossings allows to identify
involved single-particle states and quasiparticle configurations
along the yrast line (see Refs.~\cite{Voigt1983_RMP55-949, Szymanski1983_book}).

As compared with even-even nuclei, the odd-$A$ systems at low
to medium spins provide much richer experimental data and thus give
deeper insight into single-particle and shell structures in the vicinity of
the Fermi level.  This can be illustrated by the Tm $(Z=69)$ isotopes
which are used in the present manuscript as a testing ground for
different theoretical approaches. For example, the ground state bands
(GSB) in the $N=96-102$ Tm isotopes are build on the $\pi1/2^+[411]$
Nilsson state. Recently, the experimental evidence of a sharp backbending
in this band has been observed in the $^{169}$Tm nucleus~\cite{Asgar2017_PRC95-031304R}.
The backbending is also sharp in this band in $^{165}$Tm~\cite{Jensen2001_NPA695-3}.
On the contrary, the situation is completely different in the $^{167}$Tm nucleus
(which is located between two above mentioned nuclei) the GSB
of  which shows only a smooth upbending ~\cite{Burns2005_JPG31-S1827}.
These features could provide a detailed information on the single-particle level
distribution in the vicinity of the Fermi level and
yrast-yrare interaction~\cite{Bengtsson1978_PLB73-259, Wu1991_PRL66-1022}.

Over the years the experimental data on rotating rare-earth nuclei have
been used  as a testing ground for various theoretical models such as the
cranked Nilsson-Strutinsky method~\cite{Andersson1976_NPA268-205},
the cranked shell model (CSM) with Nilsson~\cite{Bengtsson1979_NPA327-139}
and Woods-Saxon~\cite{Nazarewicz1985_NPA435-397, Cwiok1987_CPC46-379} potentials,
the projected shell model~\cite{Hara1995_IJMPE4-637}, the tilted axis
cranking model~\cite{Frauendorf2001_RMP73-463}, the cranked relativistic
(covariant)~\cite{Afanasjev1996_NPA608-107, Afanasjev2000_PRC62-054306}
and non-relativistic density functional theories (DFTs)~\cite{Terasaki1995_NPA593-1,
Egido1993_PRL70-2876, Afanasjev2000_PRC62-054306}, etc.
They differ by employed assumptions and approximations and range from simple
CSM based on phenomenological potentials to much more microscopic
cranked DFTs.  In the present manuscript, the experimental data on the $N=96-102$
Er $(Z=68)$, Tm $(Z=69)$ and Yb $(Z=70$) nuclei will be used for a comparative analysis of three
different theoretical approaches, namely,
\begin{enumerate}[(i)]
  \item The cranked relativistic Hartree-Bogoliubov approach with pairing correlations
  treated by approximate particle number projection by means of  the Lipkin-Nogami
  method (further abbreviated as CRHB+LN \cite{Afanasjev2000_NPA676-196}).
  \item The cranking covariant density functional theory with pairing correlations
  treated by the shell-model-like approach (further abbreviated as cranking CDFT-SLAP
  \cite{Shi2018_PRC97-034317}).
  \item The particle-number conserving method based on the cranked shell model
  in which the phenomenological Nilsson potential is adopted for the mean field
  (further abbreviated as PNC-CSM~\cite{Zeng1994_PRC50-1388}).
\end{enumerate}
The first two methods are based on covariant density functional theory (CDFT), while the
latter one on phenomenological Nilsson potential.  The latter two approaches use the
same particle-number conserving method, while the first one is based on approximate
particle number projection by means of the Lipkin-Nogami (LN) method.
The goals of this study are (i) to evaluate the weak and strong points of these approaches,
(ii) to estimate to which extent approximate particle number projection by means of the
LN method is a good approximation to the particle-number conserving method, and
(iii) to evaluate typical accuracy of the description of experimental data
by these methods.

CDFT~\cite{Ring1996_PPNP37-193, Vretenar2005_PR409-101, Meng2006_PPNP57-470}
is well suited for the description of rotational structures.
It exploits basic properties of QCD at low energies, in particular,
the symmetries and the separation of scales~\cite{Lalazissis2004_book}.
Built on the Dirac equation, it provides a consistent treatment of the
spin degrees of freedom~\cite{Lalazissis2004_book, Cohen1992_PRC45-1881}
and spin-orbit splittings~\cite{Bender1999_PRC60-034304, Litvinova2011_PRC84-014305}.
The latter has significant influence on the shell structure.
It also includes the complicated interplay between the
large Lorentz scalar and vector self-energies induced on the QCD
level by the in-medium changes of the scalar and vector quark
condensates~\cite{Cohen1992_PRC45-1881}.
Lorentz covariance of CDFT leads to the fact
that time-odd mean fields of this theory are determined as spatial
components of Lorentz vectors and therefore coupled with the same
constants as the time-like components~\cite{Afanasjev2010_PRC81-014309},
which are fitted to ground-state properties of finite nuclei.
This is extremely important for the description of nuclear
rotations~\cite{Afanasjev2000_PRC62-031302R, Vretenar2005_PR409-101, Meng2013_FP8-55}.
Using cranked versions of CDFT, many rotational phenomena
such as superdeformation at high spin~\cite{Konig1993_PRL71-3079, Afanasjev1996_NPA608-107, Afanasjev2000_NPA676-196},
smooth band termination~\cite{Vretenar2005_PR409-101},
magnetic~\cite{Madokoro2000_PRC62-061301R, Peng2008_PRC78-024313, Zhao2011_PLB699-181, Zhao2015_PRC92-034319}
and antimagnetic~\cite{Zhao2011_PRL107-122501, Zhao2012_PRC85-054310, Liu2019_PRC99-024317} rotations,
nuclear chirality~\cite{Zhao2017_PLB773-1},
clusterization at high spins~\cite{Zhao2015_PRL115-022501,
Zhao2018_JEMPE27-1830007, Afanasjev2018_PRC97-024329, Ren2019_SCPMA62-112062},
the birth and death of particle-bound rotational bands and the extension of nuclear landscape
beyond spin zero neutron drip line~\cite{Afanasjev2019_PLB794-7} have been investigated successfully.

The Nilsson potential~\cite{Nilsson1955_DMFM29-16, Nilsson1969_NPA131-1, Nilsson1995_Book}
has been used in the calculations of rotational properties for more that half of century.
Contrary to the CDFT, cranking approaches based on this potential lack full self-consistency
and do not include time-odd mean fields.
Despite that they are still quite powerful theoretical tools which have
high predictive power.
They have been instrumental in the prediction of superdeformation
and smooth band termination at high spin as well as magnetic, antimagnetic and chiral rotations
(see Refs.~\cite{Nilsson1995_Book, Afanasjev1999_PR322-1, Frauendorf2001_RMP73-463}
and references quoted therein). They are still extensively used by Lund and Notre-Dame
groups in the interpretation  of recent experimental
data~\cite{Petrache2019_PRC99-041301R, Bhattacharya2019_PRC100-014315}.
This is in part due to the fact that cranking approaches based
on the Nilsson potential are by the orders of
magnitude numerically less time consuming that those based on the DFT approaches.

Pairing correlations are extremely important for the description
of rotational properties such as the moment of inertia (MOI), the
frequencies of paired band crossings leading either to backbendings
or upbendings, the alignment gains at paired band crossings,
etc~\cite{Bohr1975_Book, Ring2004_Book, Meng2015_book}.
They are usually treated by the Bardeen-Cooper-Schrieffer (BCS)
or Hartree-Fock-Bogoliubov (HFB) approaches
within the mean-field approximation~\cite{Ring2004_Book}.
However, in these two standard methods pairing collapse takes
place either at a critical rotational frequency $\omega_{\rm c}$~\cite{Mottelson1960_PRL5-511}
or a critical temperature $T_{\rm c}$~\cite{Sano1972_PTP48-160}.
To restore this broken symmetry, a number of approximate methods of
particle number restoration have been developed in the past. One of most
widely used is the Lipkin-Nogami method~\cite{Lipkin1960_AoP9-272,
Lipkin1961_AoP12-452, Nogami1964_PR0134-B313}, which considers the second-order correction
of the particle-number fluctuation.
When this method is implemented, the pairing
collapse does not appear in the solutions of the cranked HFB equations for a
substantially large frequency range~\cite{Gall1994_ZPA348-183,
Valor1996_PRC53-172, Afanasjev2000_NPA676-196}.
In particular, the CRHB+LN calculations successfully describe
experimental data on rotational properties across the nuclear chart
and different physical phenomena such as the rotation of normal deformed nuclei,
super- and hyperdeformation at high spin, pairing  phase transition, role of
proton-neutron pairing in $N\approx Z$ nuclei,
etc~\cite{Afanasjev1999_PRC60-051303, Afanasjev2000_NPA676-196, OLeary2003_PRCC67-021301,
Afanasjev2013_PRC88-014320, Afanasjev2014_PS89-054001, Jeppesen2009_PRC80-034324,
Herzberg2009_EPJA42-333, Afanasjev2005_PRC71-064318,
Ray2016_PRC94-014310, Andreoiu2007_PRC75-041301R, Davies2007_PRC75-011302R}.
However, the investigations have shown that the LN method breaks down in the
weak pairing limit~\cite{Dobaczewski1993_PRC47-2418, Sheikh2002_PRC66-044318}
leading to pairing collapse.
This is especially a case for extremely high rotational frequencies and for rotational
bands built on multi-qp pair-broken excited configurations.
It turns out that for such situations the calculations without pairing provide quite
accurate description of experimental rotational properties~\cite{Afanasjev1996_NPA608-107,
Afanasjev1999_PR322-1, Afanasjev2005_PRC71-064318}.

In addition, various particle-number projection approaches based on the BCS or
HFB formalism have been developed over the time~\cite{Ring2004_Book,
Egido1982_NPA388-19, Anguiano2001_NPA696-467, Volya2001_PLB509-37,
Stoitsov2007_PRC76-014308, Bender2009_PRC79-044319, Yao2014_PRC89-054306}.
In these approaches, the ideal treatment is the variation
after projection.  However, such methods are very complicated and computationally
extremely expensive for deformed rotational structures.
To overcome these problems, alternative non-variational methods
aiming at the diagonalization of the many-body Hamiltonian directly have
been developed~\cite{Zeng1983_NPA405-1, Pillet2002_NPA697-141}.
In this so-called shell-model-like approach (SLAP)~\cite{Meng2006_FPC1-38},
or originally referred as particle-number conserving (PNC) method~\cite{Zeng1983_NPA405-1},
the pairing Hamiltonian is diagonalized directly in a properly truncated
Fock-space~\cite{Wu1989_PRC39-666}. In the SLAP/PNC approach,
both particle number conservation and the Pauli blocking effects are treated
exactly.  Note that the SLAP/PNC method has been built into theoretical
approaches based on CSM with the Nilsson~\cite{Zeng1994_PRC50-1388}
and Woods-Saxon~\cite{Molique1997_PRC56-1795,Fu2013_PRC87-044319}
potentials as well as on those based on relativistic~\cite{Meng2006_FPC1-38}
and non-relativistic~\cite{Liang2015_PRC92-064325} DFTs.
These methods have been successful in the description of different nuclear phenomena
in rotating nuclei such as odd-even differences in MOI~\cite{Zeng1994_PRC50-746},
identical bands~\cite{Zeng2001_PRC63-024305, Liu2002_PRC66-024320},
nuclear pairing phase transition~\cite{Wu2011_PRC83-034323},
antimagnetic rotation~\cite{Zhang2013_PRC87-054314, Zhang2016_PRC94-034305,
Liu2019_PRC99-024317, Zhang2019_ChinPhysC43-054107},
and high-$K$ rotational bands in the rare-earth nuclei~\cite{Zhang2009_PRC80-034313,
Zhang2018_PRC98-034304, He2018_PRC98-064314, Liu2019_PRC100-064307, Dai2019_CSB64},
and rotational bands in actinides~\cite{He2009_NPA817-45, Zhang2011_PRC83-011304R,
Zhang2012_PRC85-014324, Zhang2013_PRC87-054308, Fu2014_PRC89-054301}.
Note that similar approaches to treat pairing correlations with
exactly conserved particle number can be found in
Refs.~\cite{Richardson1964_NP52-221, Pan1998_PLB422-1, Volya2001_PLB509-37,
Jia2013_PRC88-044303, Jia2013_PRC88-064321, Chen2014_PRC89-014321}.

The paper is organized as follows. Theoretical frameworks of the CRHB+LN,
cranking CDFT-SLAP and PNC-CSM approaches are presented in Sec.~\ref{Sec:Theor}.
The structure of point-coupling and meson-exchange
covariant energy density functionals (CEDFs) and of the
Nilsson potential is considered in this section too. Two methods for the treatment
of pairing, i.e., the SLAP (or PNC) and LN,  are also discussed.
The numerical details of the present calculations are given in Sec.~\ref{Sec:Numerical}.
The results of the calculations for even-even Er and Yb isotopes obtained within
these three approaches as well as a detailed comparison of these results are
reported in Sec.~\ref{Sec:CDFT_results}.
The results for odd-proton Tm nuclei are presented in Sec.~\ref{Sec:PNC}; because
of numerical limitations the major focus is  on the excitation energies and MOIs of the
1- and 3-qp configurations obtained in the PNC-CSM calculations. In addition, the evolution
of backbendings/upbendings with increasing neutron number is discussed.
 Finally, Sec.~\ref{Sec:summary} summarizes the results
of our work.

\section{Theoretical framework \label{Sec:Theor}}

In this section we will give a brief introduction to the cranked CDFT and CSM
approaches and the methods for treating the pairing correlations, namely,
SLAP and the LN  method. Note that the cranking methods discussed
are based on one-dimensional cranking approximation.

\subsection{The shell-model-like approach}

The cranking many-body Hamiltonian with pairing correlations can be written as
\begin{equation}\label{hami-eq-crank}
   \hat H=\hat H_0+\hat H_{\rm P} .
\end{equation}
Here the one-body  Hamiltonian is given by
\begin{eqnarray}
\hat H_0=\sum\hat h_0=\sum (h_{\rm s.p.}-\omega_x j_x) ,
\label{one-body-H}
\end{eqnarray}
and $\hat H_{\rm P}$ is pairing Hamiltonian.
$h_{\rm s.p.}$ and $-\omega_x j_x$ are the single-particle Hamiltonian
and Coriolis term, respectively. $h_{\rm s.p.}$ can be represented by any
mean field Hamiltonian. So far
the SLAPs based on phenomenological
Nilsson~\cite{Zeng1994_PRC50-1388} and
Woods-Saxon~\cite{Fu2013_PRC87-044319}
potentials and non-relativistic (Skyrme Hartree-Fock
approach~\cite{Liang2015_PRC92-064325})
and relativistic (CDFT~\cite{Shi2018_PRC97-034317})
DFTs have been developed.
In the present work, we employ two SLAPs: one is
based on microscopic cranked CDFT approach and another on
phenomenological cranked Nilsson Hamiltonian.

  The basic idea of SLAP is to diagonalize the many-body Hamiltonian~(\ref{hami-eq-crank}) directly
in a sufficiently large many-particle configuration
(MPC) space, characterized by an exact particle number~\cite{Zeng1983_NPA405-1},
which is constructed from the cranked single-particle states.
After diagonalizing the one-body Hamiltonian $\hat H_0$,
one can obtain the single-particle routhians $\varepsilon_{\mu\alpha}$
\begin{equation}
    \hat H_0 =\sum_{\mu\alpha} \varepsilon_{\mu\alpha}\hat b^\dag_{\mu\alpha}\hat b_{\mu\alpha} ,
\end{equation}
and the corresponding eigenstate
$|\mu\alpha\rangle$ (denoted further by $|\mu\rangle$)
\begin{equation}
    |\mu\alpha\rangle = \sum_\xi C_{\mu\xi}(\alpha)|\xi\alpha\rangle ,
\end{equation}
for each level $\mu$ with the signature $\alpha$.
Therefore, the MPC $|i\rangle$  for the $n$-particle
system can be constructed as~\cite{Zeng1994_PRC50-1388}
\begin{equation}
   |i\rangle= |\mu_1\mu_2\cdots\mu_n\rangle
            = \hat b^\dag_{\mu_1}\hat b^\dag_{\mu_2}\cdots \hat b^\dag_{\mu_n}|0\rangle .
\end{equation}
The parity $\pi$, signature $\alpha$, and the corresponding configuration energy for each
MPC are obtained from the occupied single-particle states.
By diagonalizing the cranking many-body Hamiltonian~(\ref{hami-eq-crank}) in a sufficiently
large MPC space (a dimension of 1000 for both protons and neutrons is good enough for
rare-earth nuclei),
reasonably accurate solutions for the ground state and
low-lying excited eigenstates can be obtained. Their wavefunctions
can be written as
\begin{equation}
 |\Psi\rangle = \sum_{i} C_i \left| i \right\rangle ,
 \qquad (C_i \; \textrm{real}) , \label{eq:psi}
\end{equation}
where $C_i$ are the corresponding expansion coefficients.

For the state $| \Psi \rangle$, the angular momentum alignment  is given by
\begin{equation}
\langle \Psi | J_x | \Psi \rangle = \sum_i C_i^2 \langle i | J_x | i
\rangle + 2\sum_{i<j}C_i C_j \langle i | J_x | j \rangle ,
\end{equation}
and the kinematic MOI by
\begin{equation}
J^{(1)}=\frac{1}{\omega_x} \langle\Psi | J_x | \Psi \rangle .
\end{equation}
Because $J_x$ is a one-body operator, the matrix element $\langle i | J_x | j \rangle$
($i\neq j$)  may be non-zero  only when the states
$|i\rangle$ and $|j\rangle$ differ by
one particle occupation~\cite{Zeng1994_PRC50-1388}.
After a certain permutation of creation operators,
$|i\rangle$ and $|j\rangle$ can be recast into
\begin{equation}
 |i\rangle=(-1)^{M_{i\mu}}|\mu\cdots \rangle , \qquad
 |j\rangle=(-1)^{M_{j\nu}}|\nu\cdots \rangle ,
\end{equation}
where $\mu$ and $\nu$ denote two different single-particle states,
and $(-1)^{M_{i\mu}}=\pm1$, $(-1)^{M_{j\nu}}=\pm1$
depend on whether the permutation is even or odd.
Therefore, the angular momentum alignment of
$|\Psi\rangle$ can be expressed as
\begin{equation}
 \langle \Psi | J_x | \Psi \rangle =
 \sum_{\mu} j_x(\mu) + \sum_{\mu<\nu} j_x(\mu\nu) ,
 \label{eq:jx}
\end{equation}
where the diagonal term $j_x(\mu)$ and the
off-diagonal (interference) term $j_x(\mu\nu)$  can be
written as
\begin{eqnarray}
j_x(\mu)&=&\langle\mu|j_{x}|\mu\rangle n_{\mu} , \\
j_x(\mu\nu)&=&2\langle\mu|j_{x}|\nu\rangle\sum_{i<j}(-1)^{M_{i\mu}+M_{j\nu}}C_{i}C_{j}
  \quad  (\mu\neq\nu) . \nonumber \\
\end{eqnarray}
The occupation probability $n_{\mu}$ of cranked single-particle
orbital $|\mu\rangle$ is given by
\begin{equation} \label{eq:occupation}
n_{\mu}=\sum_{i}|C_{i}|^{2}P_{i\mu} .
\end{equation}
$P_{i\mu}=1$ if $|\mu\rangle$ is occupied in MPC $|i\rangle$, and
$P_{i\mu}=0$ otherwise.  Note that in the cranking CDFT-SLAP, the occupation
probabilities will be iterated back into the densities and currents in Table~\ref{tab:dens+cur}
to achieve self-consistency~\cite{Meng2006_FPC1-38, Shi2018_PRC97-034317}.

 In general, the pairing Hamiltonian $\hat H_{\rm P}$ can be written as
\begin{equation}\label{eq-hpair-common}
    \hat H_{\rm P} = \hat H_{\rm pair-mon} + \hat H_{\rm pair-quad} + O({\rm higher\,\, order}) ,
\end{equation}
with
\begin{eqnarray}\label{eq-hpair-common}
    \hat H_{\rm pair-mon} & = & -G_0\sum_{\xi,\eta>0}^{\xi\neq\eta}\hat\beta^\dag_{\xi}\hat\beta^\dag_{\bar\xi}
                        \hat\beta_{\bar\eta}\hat\beta_{\eta} , \\
 H_{\rm pair-quad} & = &  -G_{2} \sum_{\xi\eta} q_{2}(\xi)q_{2}(\eta)
                \hat\beta^\dag_{\xi}\hat\beta^\dag_{\bar\xi}
                \hat\beta_{\bar\eta}\hat\beta_{\eta} ,
                \label{eq-quad-pair}
                \end{eqnarray}
being the Hamiltonians
of monopole and quadrupole pairing and  $G_0$ and $G_2$  their
effective pairing strengths. Higher order terms are usually neglected.
Note that $\bar\xi$ ($\bar\eta$) labels the
time-reversal state of $\xi$ ($\eta$), and $\xi\neq\eta$ means that the
self-scattering of the nucleon pairs is forbidden~\cite{Meng2006_FPC1-38}.
In Eq.\ (\ref{eq-quad-pair}), $q_{2}(\xi)$ and $q_{2}(\eta)$ are the diagonal
elements of the stretched quadrupole operator.
It turns out that reasonable agreement with experiment is obtained in cranking CDFT-SLAP
with only monopole pairing~\cite{Shi2018_PRC97-034317};
recent investigation of Ref.~\cite{Xiong2019_arxiv1908.03561} has shown that
with renormalized pairing strengths the cranking CDFT-SLAP results with monopole
pairing are quite similar to those obtained  with the separable pairing force
of Ref.~\cite{Tian2009_PLB676-44}.
Thus, we only include monopole pairing in the cranking CDFT-SLAP code.
On the contrary, the addition of quadrupole
pairing is necessary in the SLAP with Nilsson potential.

In the SLAP, the pairing energy $E_{\rm pair}$ due to pairing correlations is defined as
\begin{equation}
E_{\rm pair} = \langle\Psi|\hat H_{\rm P}|\Psi\rangle .
\label{Epair-CDFT-SLAP}
\end{equation}

\subsection{Cranked relativistic Hartree-Bogliuobov approach with approximate
particle number projections by means of Lipkin-Nogami method}
\label{CRHB+LN-sect}

The cranked relativistic Hartree-Bogoluibov (CRHB) equations with approximate particle
number projection by means of the Lipkin-Nogami (LN) method (further CRHB+LN) are given
by~\cite{Afanasjev1999_PRC60-051303, Afanasjev2000_NPA676-196}
\begin{widetext}
\begin{equation}
\left(\begin{array}{cc}
\hat{h}_{\rm D}\left(\eta\right)-\lambda\left(\eta\right)-\omega_{x}\hat{J}_{x} & \hat{\Delta}\left(\eta\right)\\
-\hat{\Delta}^{\ast}\left(\eta\right) & -\hat{h}_{\rm D}^{\ast}\left(\eta\right)+\lambda\left(\eta\right)+\omega_{x}\hat{J}_{x}^{\ast}
\end{array}\right)\left(\begin{array}{c}
U\left(\bm{r}\right)\\
V\left(\bm{r}\right)
\end{array}\right)_{k}=E_{k}\left(\eta\right)\left(\begin{array}{c}
U\left(\bm{r}\right)\\
V\left(\bm{r}\right)
\end{array}\right)_{k} ,
\end{equation}
\end{widetext}
where
\begin{eqnarray}
  \hat{h}_{\rm D}(\eta) &= & \hat{h}_{\rm D}  + 2 \lambda_2\,[(1+\eta)\rho - {\rm Tr}(\rho)]
  \label{h_d} ,  \\
  \hat{\Delta}(\eta) & =& \hat{\Delta} - 2\lambda_2(1-\eta)\kappa ,   \\
  \lambda(\eta) &=&  \lambda_1+\lambda_2\,[1 +\eta] , \\
  \label{equasi}
  E_k(\eta) &=&  E_k -\eta \lambda_2 .
\end{eqnarray}
Here $\hat{h}_{\rm D}$ is the single-nucleon Dirac Hamiltonian
the structure of which is discussed in more detail in Sec.~\ref{CEDFs-classes}.
$\hat{\Delta}$ is the pairing potential, $U_k$ and
$V_k$ are quasiparticle Dirac spinors and $E_k$ denote
the quasiparticle energies.

The $\lambda_2$ value used in the CRHB+LN calculations is given by
\begin{eqnarray}
  \lambda_2=-\frac{1}{4} \frac{{\rm Tr}_2 {\rm Tr}_2(\kappa^\ast\rho \,\overline{v}\,\sigma \kappa )}
      {[{\rm Tr}(\kappa\kappa^\dagger)]^2
        -2{\rm Tr}(\kappa\kappa^\dagger\kappa\kappa^\dagger)} .
\label{lambda2-Gog}
\end{eqnarray}
where $\sigma=1-\rho$ and
$\overline{v}_{abcd}=\langle ab | V^{pp} | cd-dc\rangle$
is antisymmetrized matrix element of the two-particle interaction
$V^{pp}$. The trace ${\rm Tr}_2$ represents the summation in the
particle-particle channel.  Note that the density matrix $\rho$ and
pairing  tensor $\kappa$ entering into Eq.\ (\ref{lambda2-Gog})
are real.

The presence of the parameter  $\eta\, (\eta=0,\pm1)$  is the
consequence of the fact that the form of the CRHB+LN equations
is not unique (see Ref.\ \cite{Afanasjev2000_NPA676-196} for
detail). The application of the LN method leads to a modification
of the CRHB equations for the fermions, while the mesonic part of
the CRHB theory is not affected. This modification is obtained by
the restricted variation of $\lambda_2 \langle (\Delta N)^2 \rangle$,
namely, $\lambda_2$ is not varied and its value is calculated
self-consistently using Eq.\ (\ref{lambda2-Gog}) in each step of
the iteration.  In the present calculations we are using the case of
$\eta=+1$ which provides reasonable numerical stability of
the CRHB+LN equations. It corresponds to the shift of whole
modification into  the particle-hole channel of the CRHB+LN theory:
$\hat{h}_{\rm D} \rightarrow \hat{h}_{\rm D} + 4\lambda_2 \rho -2 \lambda_2  {\rm Tr}(\rho)$
leaving  pairing potential $\hat{\Delta}$ unchanged.

In the CRHB theory the phenomelogical Gogny-type finite range interaction
\begin{eqnarray}
V^{pp}(1,2) = && f \sum_{i=1,2} e^{-[({\bm r}_1-{\bm r} _2)/\mu_i]^2} \nonumber\\
              &&\times (W_i+B_i P^{\sigma}- H_i P^{\tau} - M_i P^{\sigma} P^{\tau}) ,
\label{Vpp}
\end{eqnarray}
is used in the pairing channel. Here $\mu_i$, $W_i$, $B_i$, $H_i$ and
$M_i$ $(i=1,2)$ are the parameters of the force and $P^{\sigma}$ and $P^{\tau}$
are the exchange operators for the spin and isospin variables, respectively.
The parameter set D1S~\cite{Berger1991_CPC63-365} is employed for the
Gogny pairing force. A scaling factor $f$ is used here for
fine tuning of pairing properties to the mass region under study
\cite{Afanasjev2013_PRC88-014320}.
A clear advantage of the Gogny pairing force is that all
multipoles of the interaction are taken into account in the pairing channel.

The expectation value of the total  angular momentum along the
rotational axis is given by
\begin{eqnarray}
J={\rm Tr} (j_x \rho),
\end{eqnarray}
and the size of pairing correlations is measured in terms of the
pairing energy
\begin{eqnarray}
E_{\rm pair}=-\frac{1}{2}\rm{Tr} (\Delta\kappa).
\label{Epair}
\end{eqnarray}
This is not an experimentally accessible quantity, but
it is a measure for the size of the pairing correlations
in theoretical calculations.

\subsection{Covariant energy density functionals }
\label{CEDFs-classes}

The cranking CDFT-SLAP and CRHB+LN calculations are performed with CEDFs
representative of two classes of CDFT models~\cite{Agbemava2014_PRC89-054320},
namely, (i) those based on meson exchange with non-linear meson couplings (NLME),
and (ii) those based on point coupling (PC) models with zero-range interaction terms in the Lagrangian.
In NLME models, the exchange of mesons with finite masses leads to finite-range interaction.
In PC models, the gradient terms simulate the effects of finite range.

The Lagrangians of these two classes of the functionals can be written as:
$\mathcal{L} = \mathcal{L}_{\rm common} + \mathcal{L}_{\rm model-specific}$
where the $\mathcal{L}_{\rm common}$ consists of the Lagrangian of the
free nucleons and the electromagnetic interaction.
It is identical for all two classes of functionals and is written as
\begin{eqnarray}
\mathcal{L}_{\rm common} = \mathcal{L}^{\rm free} + \mathcal{L}^{\rm em} ,
\end{eqnarray}
with
\begin{eqnarray}
\mathcal{L}^{\rm free} = \bar{\psi}(i\gamma_{{\mu }}\partial^\mu - m ) \psi ,
\label{lagrfree}
\end{eqnarray}
and
\begin{eqnarray}
\mathcal{L}^{\rm em} = -\frac{1}{4} F^{\mu\nu} F_{\mu\nu} - e\frac{1-\tau_3}{2}\bar{\psi} \gamma^\mu\psi A_\mu .
\label{lagrem}
\end{eqnarray}

For each model there is a specific term in the Lagrangian: for the NLME models
we have
\begin{eqnarray}
\mathcal{L}_{\rm NLME} &=&  \frac{1}{2}(\partial{\sigma})^2 - \frac{1}{2} m_{\sigma}^2 \sigma^2 - \frac{1}{4}\Omega_{\mu\nu}\Omega^{\mu\nu} + \frac{1}{2} m_\omega^2\omega^2
\notag\\
&&-\frac{1}{4}\vec{R}_{\mu\nu}\vec{R}^{\mu\nu} + \frac{1}{2}m_\rho^2 \vec{\rho}^{\,2}-g_{\sigma}(\bar{\psi}\psi)\sigma
\notag\\
&&-g_\omega(\bar{\psi}\gamma_\mu\psi)\omega^\mu - g_\rho (\bar{\psi}\vec{\tau}\gamma_\mu\psi)\vec{\rho}^{\,\mu}
\notag\\
&&- \frac{1}{3} g_2 \sigma^3 - \frac{1}{4}g_3 \sigma^4 . 
\label{lagrddme}
\end{eqnarray}
Note that non-linear $\sigma$ meson couplings are important for the description
of surface properties of finite nuclei, especially the incompressibility~\cite{Boguta1977_NPA292-413}
and for nuclear deformations~\cite{Gambhir1990_AoP198-132}.
In the present manuscript, we are using NL1 \cite{Reinhard1986_ZPA323-13}
and NL5(E) \cite{Agbemava2019_PRC99-014318} CEDFs for NLME models; they depend on 6
parameters, namely, on  $m_\sigma$, $g_\sigma$, $g_\omega$, $g_\rho$,  $g_2$, and
$g_3$.

The Lagrangian of the PC models contains three parts:
\\
(i) the four-fermion point coupling terms:
\begin{eqnarray}
\begin{aligned}
\mathcal{L}^{\rm 4f} = & - \frac{1}{2}\alpha_S (\bar{\psi}\psi)(\bar{\psi}\psi) - \frac{1}{2}\alpha_V (\bar{\psi}\gamma_{\mu}\psi)(\bar{\psi}\gamma^{\mu}\psi) & \\
                  & - \frac{1}{2}\alpha_{TS} (\bar{\psi}\vec{\tau}\psi)(\bar{\psi}\vec{\tau}\psi)
                  - \frac{1}{2}\alpha_{TV} (\bar{\psi}\vec{\tau}\gamma_{\mu}\psi)(\bar{\psi}\vec{\tau}\gamma^{\mu}\psi),&
\end{aligned}
\label{lagr4f}
\end{eqnarray}
(ii) the gradient terms which are important to simulate the effects of finite range:
\begin{eqnarray}
\begin{aligned}
\mathcal{L}^{\rm der} =
& -\frac{1}{2}\delta_S \partial_{\nu}{(\bar{\psi}\psi)}\partial^{\nu}{(\bar{\psi}\psi)} & \\
& -\frac{1}{2}\delta_V \partial_{\nu}{(\bar{\psi}\gamma_{\mu}\psi)}\partial^{\nu}{(\bar{\psi}\gamma^{\mu}\psi)}& \\
& -\frac{1}{2}\delta_{TS} \partial_{\nu}{(\bar{\psi}\vec{\tau}\psi)}\partial^{\nu}{(\bar{\psi}\vec{\tau}\psi)}&\\
& -\frac{1}{2}\delta_{TV} \partial_{\nu}{(\bar{\psi}\vec{\tau}\gamma_{\mu}\psi)}\partial^{\nu}{(\bar{\psi}\vec{\tau}\gamma^{\mu}\psi)},&
\end{aligned}
\label{lagr4f}
\end{eqnarray}
(iii) The higher order terms which are responsible for the surface properties:
\begin{eqnarray}
\begin{aligned}
\mathcal{L}^{\rm hot}  =
& -\frac{1}{3}\beta_S (\bar{\psi}\psi)^3 - \frac{1}{4}\gamma_S (\bar{\psi}\psi)^4 &   \\
& -\frac{1}{4}\gamma_V[(\bar{\psi}\gamma_{\mu}\psi)(\bar{\psi}\gamma^{\mu}\psi)]^2 . &
\end{aligned}
\label{lagrhot}
\end{eqnarray}
For the PC models we have 9 parameters $\alpha_S$, $\alpha_V$, $\alpha_{TV}$, $\delta_S$, $\delta_V$,
$\delta_{TV}$, $\beta_S$, $\gamma_S$, $\gamma_V$. In these calculations we neglect the scalar-isovector
channel, i.e., we use $\alpha_{TS}=\delta_{TS}=0$, because it has been shown in Ref.~\cite{Roca-Maza2011_PRC84-054309}
that the information on masses and radii of finite nuclei does not allow to distinguish the effects of  two
isovector mesons $\delta$ and $\rho$. For PC model we are using PC-PK1 CEDF \cite{Zhao2010_PRC82-054319}.

The solution of these Lagrangians leads to the Dirac equation for the fermions and, in the
case of meson exchange models, to the Klein-Gordon equations for the mesons.
The single-particle Dirac Hamiltonian is given by
\begin{equation}\label{eq:dirac-sp}
  \hat h_{\rm D} = \bm \alpha\cdot(-i{\bm\nabla}-\bm{V})+\beta(m+S)+V^0 ,
\end{equation}
and it enters into the solutions of the cranking CDFT-SLAP [Eq.~(\ref{one-body-H}) under
the condition $\hat h_{\rm s.p.} = \hat h_{\rm D}$] and CRHB+LN
[see Eq.~(\ref{h_d})] equations.

The time-independent inhomogeneous Klein-Gordon equations for the mesonic fields obtained by means
of variational principle are given in the NLME models
as~\cite{Afanasjev1999_PRC60-051303,Afanasjev2000_NPA676-196}
\begin{eqnarray}
\left\{-\Delta-({\omega}_x\hat{l}_x)^2 + m_\sigma^2\right\}~
\sigma(\bm r) & = &
-g_\sigma \rho_{\sl S}(\bm r) \nonumber \\
\,\,\,\,\,\,\,\,\,\,\,\,\,\,\,\,\,-g_2\sigma^2(\bm r)-g_3\sigma^3(\bm r) ,
\nonumber \\
\left\{-\Delta-({\omega}_x\hat{l}_x)^2+m_\omega^2\right\}
\omega_0(\bm r)&=&
g_\omega \rho_{V}(\bm r) ,
\nonumber \\
\left\{-\Delta-[{\omega}_x(\hat{l}_x+\hat{s}_x)]^2+
m_\omega^2\right\}~
\bm\omega(\bm r)&=&
g_\omega\bm j_{V}(\bm r) ,
\nonumber \\
\left\{-\Delta-({\omega}_x\hat{l}_x)^2+m_\rho^2\right\}
\rho_0(\bm r)&=&
g_\rho\rho_{TV}(\bm r) ,
\nonumber \\
\left\{-\Delta-[{\omega}_x(\hat{l}_x+\hat{s}_x)]^2+
m_\rho^2\right\}~
\bm\rho(\bm r)&=& g_\rho \bm j_{TV}(\bm r) ,
\nonumber \\
-\Delta~A_0(\bm r)&=&e\rho_{V}^p(\bm r) ,
\nonumber \\
-\Delta~\bm A(\bm r)&=&e\bm j_{V}^p(\bm r) .
\label{KGeq}
\end{eqnarray}
No such equations are present in the PC models.

The form of the relativistic fields
$S({\bm r})$ and $V^\mu(\bm r)$ as well as the currents and densities
defining these fields depends on the class of the functional; the detailed
expressions for them are  given in Tables~\ref{tab:fields}
and~\ref{tab:dens+cur}.
 Note that so far the CRHB+LN calculations were based only on the NLME
models~\cite{Afanasjev1999_PRC60-051303,Afanasjev2000_NPA676-196, Afanasjev2003_PRC67-024309,
Afanasjev2013_PRC88-014320, Afanasjev2014_PS89-054001}.
In this manuscript, we continue to use such an
approach for a consistency with previous studies.
After solving self-consistently the equations of motion for the nucleons
[Eq.~(\ref{eq:dirac-sp})] and mesons [Eq.~(\ref{KGeq})], the total energy of the
system can be obtained; we refer the reader to Sec.~2.1. of Ref.~\cite{Afanasjev2000_NPA676-196}
for more details on this step in the CRHB+LN framework.
In the cranking CDFT-SLAP, both the NLME and PC models are used.
Note that there is no meson in the PC model,
and only the Dirac equation for the nucleons~[Eq.~(\ref{eq:dirac-sp})] exists.
The occupation probabilities $n_\mu$ of each orbital obtained
by Eq.~(\ref{eq:occupation}) will be iterated back into
the densities and currents in Table~\ref{tab:dens+cur} to achieve self-consistency
when solving the Dirac equation~\cite{Meng2006_FPC1-38, Shi2018_PRC97-034317}.

In CDFT, the quadrupole moments $Q_{20}$ and $Q_{22}$ are calculated by
\begin{eqnarray}
Q_{20} & = & \int d^{3}r\left[\rho(r)\sqrt{\frac{5}{16\pi}}\left(3z^{2}-r^{2}\right)\right] ,\\
Q_{22} & = & \int d^{3}r\left[\rho(r)\sqrt{\frac{15}{32\pi}}\left(x^{2}-y^{2}\right)\right] ,
\end{eqnarray}
and the deformation parameters $\beta$ and $\gamma$ can be extracted from
\begin{eqnarray}
\beta  =  \sqrt{a_{20}^{2}+2a_{22}^{2}} , \quad
\gamma =  \arctan\left[\sqrt{2}\frac{a_{22}}{a_{20}}\right] ,
\end{eqnarray}
using
\begin{equation}
Q_{20}=\frac{3A}{4\pi}R_{0}^{2}a_{20} , \quad
Q_{22}=\frac{3A}{4\pi}R_{0}^{2}a_{22} ,
\end{equation}
with $R_0 = 1.2A^{1/3}$~fm. Note that in this work, the sign convention
of Ref.~\cite{Ring2004_Book} is adopted for the definition of $\gamma$.

Contrary to the Nilsson potential used in the PNC-CSM approach, time-odd
mean fields emerging from space-like components of vector fields and
currents play an extremely important role in the description of rotating
nuclei in the CDFT framework \cite{Konig1993_PRL71-3079,
Afanasjev2000_PRC62-031302R, Afanasjev2010_PRC82-034329}.
They significantly affect the MOIs, single-particle alignments
and  band crossing features. Available comparisons between theory and
experiment in paired and unpaired regimes of rotation strongly suggest
that time-odd mean fields are well described by the state-of-the-art CEDFs
(see Refs.~\cite{Meng2015_book, Vretenar2005_PR409-101, Afanasjev2010_PRC82-034329}).
In contrast to non-relativistic DFTs, they are constrained by the Lorentz
covariance and thus do not require additional parameters \cite{Afanasjev2010_PRC81-014309}.

\newsavebox\boxnlme
\begin{lrbox}{\boxnlme}
\begin{minipage}{0.45\textwidth}
\begin{align*}
 S(\bm r) =& g_\sigma\sigma(\bm r) \\
 V_0(\bm r) =& g_\omega\omega_0(\bm r)+g_\rho\tau_3\rho_0(\bm r)
+e \frac{1-\tau_3} {2} A_0(\bm r) \\ ~\\
\bm V(\bm r)=&g_\omega\bm\omega(\bm r)
+g_\rho\tau_3\bm\rho(\bm r)+
e\frac {1-\tau_3} {2} \bm A(\bm r) \\
\end{align*}
\end{minipage}
\end{lrbox}
\newsavebox\boxpc
\begin{lrbox}{\boxpc}
\begin{minipage}{0.45\textwidth}
\begin{align*}
 S(\bm r)  =&  \alpha_S\rho_S+\beta_S\rho^2_S+\gamma_S\rho^3_S+\delta_S\Delta\rho_S \\
 V_0(\bm r)=&\alpha_V\rho_V+\gamma_V\rho^3_V+\delta_V\Delta\rho_V \\
               & +\tau_3\alpha_{TV}\rho_{TV}
                +\tau_3\delta_{TV}\Delta\rho_{TV}+e\frac{1-\tau_3}{2}{A}^0 \\
 \bm{V}({\bm r}) =& \alpha_V\bm{j}_V+\gamma_V(\bm{j}_V)^3+\delta_V\Delta\bm{j}_V \\
                    & +\tau_3\alpha_{TV}\bm{j}_{TV}+\tau_3\delta_{TV}\Delta\bm{j}_{TV}
                    +e\frac{1-\tau_3}{2}{\bm A}
\end{align*}
\end{minipage}
\end{lrbox}
\begin{table*}
\caption{\label{tab:fields}
 Relativistic fields $S(\bm r)$, $V_0(\bm r)$ and $\bm V(\bm r)$
as defined in non-linear meson exchange and point coupling models.}
\begin{center}
\begin{spacing}{1.2}
\def\temptablewidth{17cm}
\begin{tabular*}{\temptablewidth}{@{\extracolsep{\fill}}cc}
\hline\hline
  NLME              & PC               \\
  \hline
  \usebox{\boxnlme} & \usebox{\boxpc}  \\
\hline\hline
\end{tabular*}
\end{spacing}
\end{center}
\end{table*}

\newsavebox\boxcrhb
\begin{lrbox}{\boxcrhb}
\begin{minipage}{0.45\textwidth}
\begin{align*}
  \rho_{S}(\bm{r})        & =
   \sum_{k>0}[V_{k}^{n}(\bm{r})]^{\dagger}\hat{\beta}V_{k}^{n}(\bm{r})+
   [V_{k}^{p}(\bm{r})]^{\dagger}\hat{\beta}V_{k}^{p}(\bm{r})  \\
  \rho_{V}(\bm{r})        & =
   \sum_{k>0}[V_{k}^{n}(\bm{r})]^{\dagger}V_{k}^{n}(\bm{r})+
   [V_{k}^{p}(\bm{r})]^{\dagger}V_{k}^{p}(\bm{r})             \\
  \rho_{TV}(\bm{r})       & =
   \sum_{k>0}[V_{k}^{n}(\bm{r})]^{\dagger}V_{k}^{n}(\bm{r})-
   [V_{k}^{p}(\bm{r})]^{\dagger}V_{k}^{p}(\bm{r})             \\
  \bm{j}_{V}(\bm{r})  & =
   \sum_{k>0}[V_{k}^{n}(\bm{r})]^{\dagger}\hat{\bm{\alpha}}V_{k}^{n}(\bm{r})+
   [V_{k}^{p}(\bm{r})]^{\dagger}\hat{\bm{\alpha}}V_{k}^{p}(\bm{r})  \\
  \bm{j}_{TV}(\bm{r}) & =
   \sum_{k>0}[V_{k}^{n}(\bm{r})]^{\dagger}\hat{\bm{\alpha}}V_{k}^{n}(\bm{r})-
   [V_{k}^{p}(\bm{r})]^{\dagger}\hat{\bm{\alpha}}V_{k}^{p}(\bm{r})  \\
\end{align*}
\end{minipage}
\end{lrbox}
\newsavebox\boxslap
\begin{lrbox}{\boxslap}
\begin{minipage}{0.45\textwidth}
\begin{align*}
  \rho_{S}(\bm{r}) & =
   \sum_{\mu}n_{\mu}\bar{\psi}_{\mu}(\bm{r})\psi_{\mu}(\bm{r})                 \\
  \rho_{V}(\bm{r}) & =
   \sum_{\mu}n_{\mu}\psi_{\mu}^{\dagger}(\bm{r})\psi_{\mu}(\bm{r})              \\
  \rho_{TV}(\bm{r}) & =
   \sum_{\mu}n_{\mu}\psi_{\mu}^{\dagger}(\bm{r})\bm{\tau}_{3}\psi_{\mu}(\bm{r}) \\
  \bm{j}_{V}(\bm{r}) & =
   \sum_{\mu}n_{\mu}\psi_{\mu}^{\dagger}(\bm{r})\bm{\alpha}\psi_{\mu}(\bm{r})   \\
  \bm{j}_{TV}(\bm{r}) & =
   \sum_{\mu}n_{\mu}\psi_{\mu}^{\dagger}(\bm{r})\bm{\alpha}\tau_{3}\psi_{\mu}(\bm{r})  \\
  \rho_{c}(\bm{r}) & =
   \sum_{\mu}n_{\mu}\psi_{\mu}^{\dagger}(\bm{r})\frac{1-\tau_{3}}{2}\psi_{\mu}(\bm{r})
\end{align*}
\end{minipage}
\end{lrbox}
\begin{table*}
\caption{\label{tab:dens+cur}  Local densities and currents as defined in the CRHB+LN and cranking CDFT-SLAP.
The sums are taken over only the states with positive energies (no-sea approximation).
The indexes $n$ and $p$ indicate
neutron and proton states, respectively.
Note that the spatial
components of the electromagnetic vector potential $\bm A$ are neglected since their
contributions are extremely small.}
\begin{center}
\begin{spacing}{1.2}
\def\temptablewidth{17cm}
\begin{tabular*}{\temptablewidth}{@{\extracolsep{\fill}}cc}
\hline\hline
  CRHB+LN           & CDFT-SLAP          \\
  \hline
  \usebox{\boxcrhb} & \usebox{\boxslap}  \\
\hline\hline
\end{tabular*}
\end{spacing}
\end{center}
\end{table*}

\subsection{Cranked Nilsson model}

  The cranked Nilsson Hamiltonian is used in the PNC-CSM;
here we present a short review of its features.
The cranked shell model Hamiltonian is given by
\begin{equation}
h_0 = h_{\rm Nil} - \omega_x j_x ,
\end{equation}
 where $h_{\rm Nil}$ is the Nilsson Hamiltonian and $-\omega_x j_x$ is the
Coriolis term. Note that the collective rotation of the nucleus is considered
in the one-dimensional cranking approximation for which  the
nuclear field is rotated with the cranking frequency $\omega_x$ about
the principal $x$ axis.

The  Nilsson Hamiltonian is based on axially deformed
modified-oscillator potential
\begin{eqnarray}
V_{\rm osc}&=&\frac{1}{2} M\left[\omega_\bot^2(x^2+y^2)+\omega_z^2 z^2\right] \nonumber\\
           && + C\vec{l}\cdot\vec{s} + D \left(l^2 - \langle l^2 \rangle_N \right) ,
\end{eqnarray}
which includes spin-orbit term $\vec{l}\cdot\vec{s}$ and the
$l^2 - \langle l^2 \rangle_N$ term~\cite{Nilsson1995_Book}.
Note that the oscillator
frequencies $\omega_\bot$ and $\omega_z$ are the functions of deformation
parameters. The restriction to axial shapes is an approximation
which follows from non-selfconsistent nature of the PNC-CSM in which the
deformation of the potential is defined by the deformation parameters of the
ground state (which are axially symmetric in the region under study) and the
variations in the deformation parameters with angular momentum are neglected.

The Nilsson Hamiltonian is usually written in stretched coordinates
\begin{equation}
\xi   =  x \sqrt{\frac{M \omega_\bot}{\hbar}} , \quad
\eta  =  y \sqrt{\frac{M \omega_\bot}{\hbar}} , \quad
\zeta =  z \sqrt{\frac{M \omega_z}{\hbar}}    ,
\end{equation}
which allow to transform away the coupling terms of the
$r^2P_2(\cos\theta)$ term between the major
$N$ and $N\pm2$ shells ~\cite{Nilsson1955_DMFM29-16}.
In these coordinates the Nilsson Hamiltonian $h_{\rm Nil}$ is written
as~\cite{Nilsson1969_NPA131-1}
\begin{eqnarray}\label{eq:axialNil}
h_{\rm Nil} =&& \frac{1}{2} \hbar\omega_{0}(\varepsilon_2,\varepsilon_4)
                \left[-\nabla^{2}_\rho + \frac{1}{3}
                \left(2\frac{\partial^2}{\partial\zeta^2}-\frac{\partial^2}{\partial\xi^2}
                -\frac{\partial^2}{\partial\eta^2}\right) \right. \nonumber\\
            +&& \left.\rho^{2}-\frac{2}{3}\varepsilon_2\rho^{2}P_{2}(\cos\theta_t) +
                2\varepsilon_{4} \rho^{2}P_{4}(\cos\theta_t)\right] \nonumber\\
            -&& 2 \kappa\hbar\mathring{\omega}_{0}
                \left(\vec{s}\cdot\vec{l}_{t}-\mu(\rho^{4}-\langle\rho^{4}\rangle_{N})\right) ,
\end{eqnarray}
where $\rho^2 = \xi^2 + \eta^2 + \zeta^2$ and
$\theta_t$ ($\cos \theta_t=\zeta/\rho$) and $\vec{l}_{t}$ are the angle
and angular momentum
in the stretched coordinates, respectively. Here ($\kappa, \mu$) are
the Nilsson parameters and ($\varepsilon_2,\varepsilon_4$) are
the deformation parameters; they represent the input parameters
of the Nilsson Hamiltonian the definition of which is discussed
in Sec.~\ref{par-PNC-CSM}.
Neutron and proton oscillator parameters are given by~\cite{Nilsson1995_Book}
\begin{eqnarray}
\hbar \mathring{\omega}_{\rm n,p} =  41 A^{-1/3} \left( 1 \pm \frac{1}{3} \frac{N-Z}{A} \right) ,
\end{eqnarray}
where the plus/minus sign holds for neutrons/protons.
The quantity $\omega_0/\mathring{\omega}_0$ is determined by the volume conservation
condition
\begin{equation}
\frac{\omega_0^3}{\mathring{\omega}_0^3} =
\frac{1}{(1+\frac{1}{3}\varepsilon_2)(1-\frac{2}{3}\varepsilon_2)^{\frac{1}{2}}}
\int_{-1}^{1} \frac{ \frac{1}{2}d(\cos\theta)}
{\left(1-\frac{2}{3}\varepsilon_2 P_2 + 2 \varepsilon_4 P_4\right)^{\frac{3}{2}}} .
\end{equation}

\section{\label{Sec:Numerical}Numerical details}

\subsection{The cranking CDFT-SLAP}

In the present cranking CDFT-SLAP calculations
the point-coupling CEDF PC-PK1~\cite{Zhao2010_PRC82-054319}
is used in the particle-hole channel and the monopole
pairing interaction is employed in the particle-particle channel.
In addition, some calculations are performed with the
meson-exchange NL5(E) CEDF~\cite{Agbemava2019_PRC99-014318}
with the goal to compare their results with those obtained with PC-PK1.
In the present work, a three-dimensional harmonic oscillator
(3DHO) basis in Cartesian coordinates with good signature quantum
number~\cite{Shi2018_PRC97-034317} is adopted for solving the equation of motions
for the nucleons and mesons. The Dirac spinors are expanded into 3DHO basis with 14 major shells.
When using meson-exchange NL5(E) CEDF, 20 major shells are used for mesons.
For both protons and neutrons, the MPC truncation energies are selected to be around 8.0~MeV,
and the dimensions of the MPC space are chosen to be equal to 1000.
This provides sufficient numerical accuracy for the rare-earth nuclei.
The effective pairing strengths are equal to 1.5 MeV both for protons and neutrons;
the neutron pairing strengths are defined by fitting the experimental
odd-even mass differences in $^{166-172}$Yb, and the proton pairing strengths
are taken the same as those for neutrons.
In addition, they are also fitted to the bandhead MOIs of
$^{170}$Yb and $^{168}$Er at $\hbar\omega_x\sim0.04$~MeV.

\subsection{The CRHB+LN approach}

The CRHB+LN calculations are performed with the NL1~\cite{Reinhard1986_ZPA323-13}
and NL5(E)~\cite{Agbemava2019_PRC99-014318}
CEDFs. The latter functional provides the best global description of the ground state
properties among  the NLME functionals~\cite{Agbemava2019_PRC99-014318}.
The NL1 is the first successful CEDF
fitted more that 30 years ago. Despite that it provides quite reasonable description of
the one-quasiparticle spectra in deformed rare-earth region~\cite{Afanasjev2011_PLB706-177}
and works extremely well in the description of rotational properties of the nuclei
across the nuclear
landscape~\cite{Afanasjev1996_NPA608-107, Afanasjev1999_PRC59-3166,
Afanasjev2000_NPA676-196,Afanasjev2013_PRC88-014320}.
All fermionic and bosonic states belonging to the shells up
to $N_F=14$ and $N_B=20$ of the 3DHO basis were taken into account in the
diagonalization of the Dirac equation and the matrix inversion of the Klein-Gordon
equations, respectively. As follows from a detailed analysis
of Ref.~\cite{Afanasjev2013_PRC88-014320} this truncation of the basis provides sufficient
accuracy of the calculations.

The scaling factor $f$ of the Gogny pairing [see Eq.\ (\ref{Vpp})]
is defined at low frequency $\hbar\omega_x =0.05$ MeV by fitting the experimental MOIs
of even-even Er and Yb nuclei used in the present study. This procedure
gives the values $f=0.957$ and $f=0.950$ for the NL1 and NL5(E) functionals,
respectively.

\subsection{The PNC-CSM}
\label{par-PNC-CSM}

\begin{table}[h]
\caption{\label{tab:defEr}
  Deformation parameters ($\varepsilon_2$, $\varepsilon_4$) adopted
in the present calculations for even-even Er and Yb isotopes (see text
for details).
}
\begin{center}
\begin{spacing}{1.2}
\def\temptablewidth{8.6cm}
\begin{tabular*}{\temptablewidth}{@{\extracolsep{\fill}}ccccc}
  \hline
  \hline
  ~               & $^{164}$Er & $^{166}$Er & $^{168}$Er & $^{170}$Er \\
  \hline
  $\varepsilon_2$ & 0.258      & 0.267      & 0.273      & 0.276      \\
  $\varepsilon_4$ & 0.001      & 0.012      & 0.023      & 0.034      \\
  \hline
  ~               & $^{166}$Yb & $^{168}$Yb & $^{170}$Yb & $^{172}$Yb \\
  \hline
  $\varepsilon_2$ & 0.246      & 0.255      & 0.265      & 0.269      \\
  $\varepsilon_4$ & 0.004      & 0.014      & 0.025      & 0.036      \\
  \hline
  \hline
\end{tabular*}
\end{spacing}
\end{center}
\end{table}

The deformation parameters ($\varepsilon_2$, $\varepsilon_4$)
of even-even Er and Yb isotopes used in PNC-CSM calculations
are taken from Ref.~\cite{Bengtsson1986_ADNDT35-15}
(see Table~\ref{tab:defEr}). For deformation parameters of odd-$A$ Tm
isotopes we use an average of the deformations of neighboring
even-even Er and Yb isotopes.

\begin{table}[h]
\caption{\label{tab:ku}
The Nilsson ($\kappa_{\rm th}$, $\mu_{\rm th}$)
parameters adopted in the present calculations compared with the
parameters of Ref.\ \cite{Bengtsson1985_NPA436-14} [labelled
as ($\kappa_{\rm stand}$, $\mu_{\rm stand}$)] and Ref.\ \cite{Bengtsson1990_NPA512-124}
[labelled as $\kappa_{\rm A150}$, $\mu_{\rm A150}$)].
}
\begin{center}
\begin{spacing}{1.2}
\def\temptablewidth{8.6cm}
\begin{tabular*}{\temptablewidth}{@{\extracolsep{\fill}}ccccccc}
  \hline\hline
  $N$     & $\kappa_{\rm th}$      & $\mu_{\rm th}$
          & $\kappa_{\rm stand}$   & $\mu_{\rm stand}$
          & $\kappa_{\rm A150}$    & $\mu_{\rm A150}$ \\
  \hline
  $\pi_4$ & 0.076   & 0.57   & 0.065   & 0.57   & 0.070   & 0.50 \\
  $\pi_5$ & 0.060   & 0.57   & 0.060   & 0.65   & 0.060   & 0.55 \\
  $\nu_5$ & 0.062   & 0.43   & 0.062   & 0.43   & 0.062   & 0.43 \\
  $\nu_6$ & 0.062   & 0.40   & 0.062   & 0.34   & 0.062   & 0.40 \\
  \hline\hline
\end{tabular*}
\end{spacing}
\end{center}
\end{table}

The Nilsson parameters ($\kappa$, $\mu$) are usually obtained
by the fit of calculated single-particle levels to experimental level schemes
in the rare-earth nuclei and actinides  \cite{Nilsson1969_NPA131-1,Bengtsson1985_NPA436-14,
Bengtsson1986_ADNDT35-15}.
The Nilsson parameters employed in the present calculations
[labeled as ($\kappa_{\rm th}$, $\mu_{\rm th}$) in Table~\ref{tab:ku}]
are obtained from the parameters of Ref.~\cite{Bengtsson1990_NPA512-124}
[labeled as ($\kappa_{\rm A150}$, $\mu_{\rm A150}$) in Table~\ref{tab:ku}]
by means of some modifications in proton subsystem, namely, by
fitting the calculated proton energy levels to experimental
1-qp excitation energies in odd-$A$ Tm isotopes.
The main difference between the Nilsson parameters of Ref.~\cite{Bengtsson1990_NPA512-124}
and the ``standard'' Nilsson parameters of Ref.~\cite{Bengtsson1985_NPA436-14}
[labeled as ($\kappa_{\rm stand}$, $\mu_{\rm stand}$)  in Table~\ref{tab:ku}]
is that the proton $Z = 64$ gap is increased and the
neutron $i_{13/2}$ orbitals are lowered by about 0.5~MeV.
This  can improve the description of the backbendings in this mass region.
These three parameter sets are shown at Table~\ref{tab:ku}.
The comparison between experimental band-head energies of the
1- and 3-qp states in odd-$A$ Tm isotopes and their calculated
counterparts obtained with these three sets of the Nilsson parameters is discussed
in Sec.~\ref{Sec:PNC}.

In the present PNC-CSM calculations, the MPC space is constructed
from proton $N=4, 5$ and neutron $N=5, 6$ major shells.
The MPC truncation energies are selected to be around
6.0~MeV for protons and 5.5~MeV for neutrons, respectively.
The dimensions of the MPC space are equal to 1000 both for protons and neutrons;
this is equivalent to the MPC space used in the cranking CDFT-SLAP calculations.
In the PNC-CSM calculations, both monopole and quadrupole pairings are considered.
The pairing strengths are defined by fitting the odd-even mass differences and
the MOIs of low-spin parts of experimental bands in even-even and odd-$A$ nuclei.
The monopole proton pairing strengths are the same for all even-even
Er and Yb isotopes and they are equal to $G_{\rm 0p}=0.35$~MeV.
On the contrary, there is some variation of the monopole neutron pairing strengths
with neutron number;  they are equal to $G_{\rm 0n}=0.40$~MeV for $N=96$ and 98 isotopes
and $G_{\rm 0n}=0.25$~MeV for $N=100$ and 102 isotopes.
For odd-$A$ nuclei $^{165,167,169,171}$Tm, the monopole pairing strengths are
$G_{\rm 0p}=0.31$~MeV and $G_{\rm 0n}=0.33$~MeV.

Previous investigations have shown that the description of experimental
bandhead energies and level crossing frequencies can be improved when
quadrupole pairing is taken into account~\cite{Diebel1984_NPA419-221,Satula1994_PRC50-2888}.
However, an accurate determination of the quadrupole pairing strength still
remains not fully solved problem.
Quadrupole pairing strengths are typically fitted in the frameworks of different models
to the bandhead energies, MOIs, and bandcrossing frequencies, and they are usually chosen to be proportional
to the strengths of monopole pairing~\cite{Diebel1984_NPA419-221, Hara1995_IJMPE4-637}.
However, the proportionality constants depend on nuclear mass region.
It was argued in Ref.~\cite{Sakamoto1990_PLB245-321} that the quadrupole pairing strength
is expected to be determined by the restoration of the Galilean invariance broken
by the monopole pairing.
However, further modifications of its strength are still needed to describe
experimental MOIs and bandcrossing frequencies in many cases.
We tried to keep quadrupole pairing strengths $G_2$ proportional to monopole
pairing strengths $G_0$ in the PNC-CSM calculations but found that resultant
small change of quadrupole pairing strength with particle number has
little influence on the calculated MOIs.
Thus, it was decided for all nuclei under study to keep the strength of quadrupole pairing
at fixed values of $G_{\rm 2p}=G_{\rm 2n}=0.006$~MeV.

\section{\label{Sec:CDFT_results}Comparison between the CRHB+LN, cranking CDFT-SLAP,
and PNC-CSM calculations for even-even Er and Yb isotopes}

\begin{figure*}[!]
\centering
\includegraphics[width=0.95\textwidth]{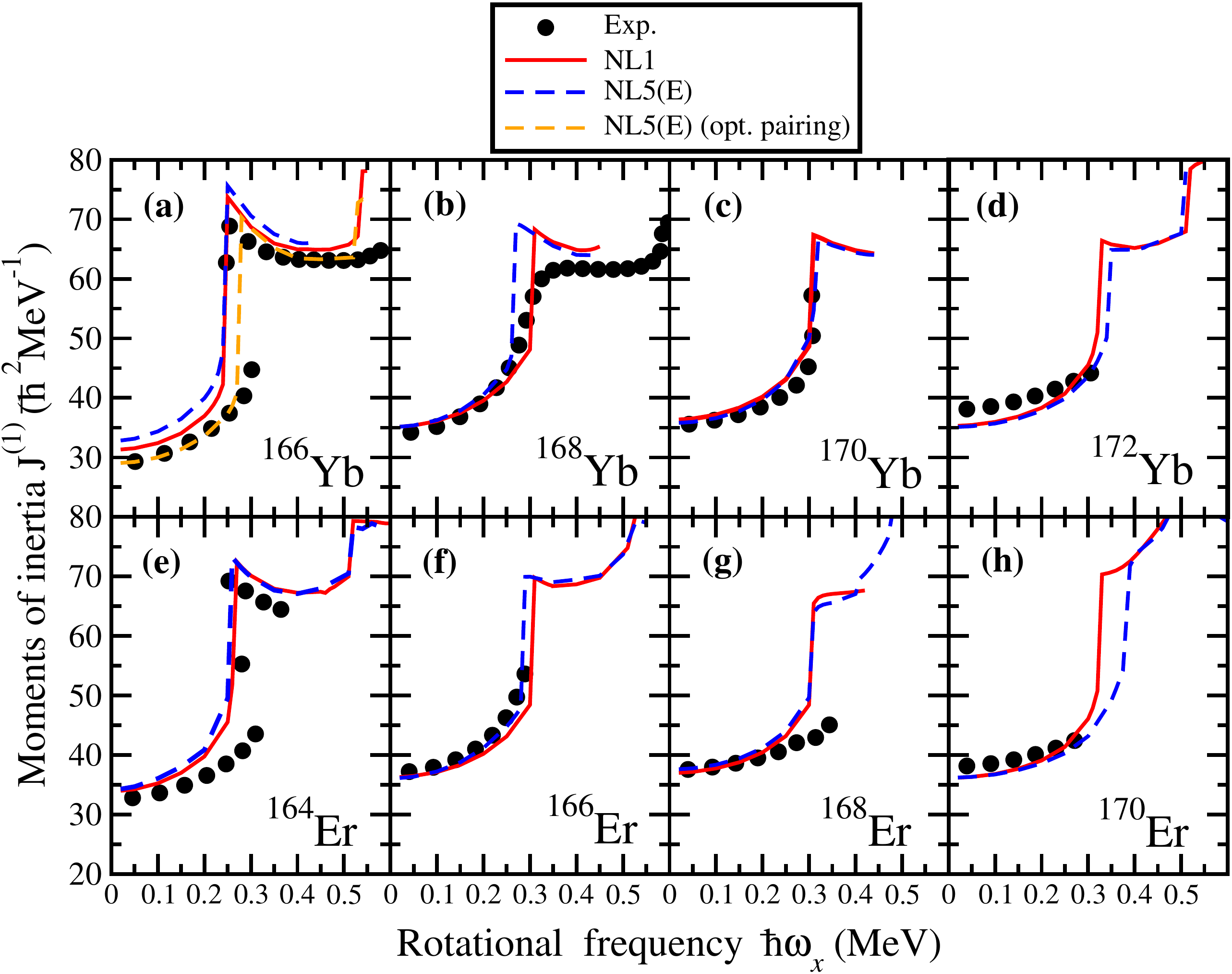}
\caption{\label{fig:CRHB-Tm}
The kinematic MOIs $J^{(1)}$ for the GSBs of
$^{166,168,170,172}$Yb (upper panels) and $^{164,166,168,170}$Er (lower panels),
obtained in the CRHB+LN calculations with the CEDFs NL1~\cite{Reinhard1986_ZPA323-13}
(red solid lines) and NL5(E)~\cite{Agbemava2019_PRC99-014318} (blue dashed lines),
compared to the experimental data (black solid circles).
The latter are taken from Refs.~\cite{Singh2018_NDS147-1, Baglin2008_NDS109-1103,
Baglin2010_NDS111-1807, Baglin2018_NDS153-1, Firestone1999}.
Orange dashed line in panel (a) shows the results obtained with the NL5(E) CEDF and
scaling factor $f$ of Gogny pairing fitted in such a way that it exactly reproduces
experimental MOI of $^{166}$Yb GSB at $\hbar\omega_x=0.05$~MeV.
}
\end{figure*}

Figure~\ref{fig:CRHB-Tm} compares experimental and theoretical
MOIs obtained in the CRHB+LN calculations with the CEDFs NL1 and NL5(E).
It can be seen that the results of the calculations for both functionals
are close to each other.  They also reproduce the experimental MOIs quite well.
Note that the CRHB+LN, cranking CDFT-SLAP and PNC-CSM calculations are
performed as a function of rotational frequency. Thus, they cannot
predict or describe the back-sloping part of the backbending curve. However,
they can reproduce an average frequency of backbending  defined as
 $\hbar\omega_{\rm ave}= \frac{1}{2}(\hbar\omega_1+\hbar\omega_2)$ where
$\hbar \omega_1$ corresponds to the frequency at which the MOI curve
bends backward and $\hbar \omega_2$ to the frequency at which the MOI
curve bends forward.

There is only a small difference in the band crossing frequencies for the
$N=98$ ($^{166}$Er and $^{168}$Yb) and $N=102$ ($^{170}$Er and $^{172}$Yb)
isotones obtained in the calculations with the NL1 and NL5(E)  functionals (see Fig.~\ref{fig:CRHB-Tm}).
 In the $N=98$ isotones [Figs.~\ref{fig:CRHB-Tm}(b) and (f)], the calculated first
upbending takes place at somewhat lower frequencies for NL5(E) as compared  with NL1.
On the contrary, the situation is reversed in the $N=102$ isotones
[Figs.~\ref{fig:CRHB-Tm}(d) and (h)]. In the $N=96$ ($^{164}$Er and $^{166}$Yb)
isotones, the first band crossing is calculated at $\hbar \omega_x \approx 0.25$~MeV.
The calculated first band crossing frequency gradually increases with increasing
neutron number and it reaches $\hbar \omega_x \approx 0.32$~MeV in the
$N=102$ ($^{170}$Er and $^{172}$Yb) isotones.
The calculated first upbendings are
very sharp in the CRHB+LN calculations for both functionals.
Experimental data show
that sharp backbendings exist in $^{166,170}$Yb [Figs.~\ref{fig:CRHB-Tm}(a) and (c)]
and $^{164}$Er  [Fig.~\ref{fig:CRHB-Tm}(e)], while the upbendings in $^{168}$Yb
[Fig.~\ref{fig:CRHB-Tm}(b)]  and $^{166}$Er [Fig.~\ref{fig:CRHB-Tm}(f)] are somewhat
smoother as  compared with calculations. Note that for the  $^{170,172}$Yb
[Figs.~\ref{fig:CRHB-Tm}(b) and (d)] and $^{166,168,170}$Er
[Figs.~\ref{fig:CRHB-Tm}(f), (g) and (h)] nuclei the $s$-bands have not been observed
experimentally.
Therefore, further experiments are needed to verify the predicted
upbending features of these nuclei.

It can be seen in Fig.~\ref{fig:CRHB-Tm}(b) that a second upbending
in $^{168}$Yb is observed experimentally at $\hbar\omega_x\sim0.58$~MeV. In
this nucleus, the CRHB+LN calculations for both functionals do not converge
above $\hbar \omega_x =0.45$~MeV. This numerical instability is most likely
caused by the competition of different configurations located at comparable
energies in the region of second band crossing. Indeed, the CRHB+LN
calculations provide converged solutions at frequencies $\hbar \omega_x \sim 0.6$
MeV in most of the nuclei under consideration even if the pairing is extremely weak.
These solutions are not shown in Fig.\ \ref{fig:CRHB-Tm} if there
is non-convergence in the region of second band crossing. Thus, this non-convergence
should not  necessary be a manifestation of
the deficiencies of the LN method. Note that similar situation with non-convergence
of the CRHB+LN solutions in the region of second band crossing has been
observed also in rotational structures of some actinides and light superheavy
nuclei (see Ref.~\cite{Afanasjev2013_PRC88-014320}).

Note that the CRHB+LN calculations converge in most of even-even
nuclei. For example, they predict second upbending in $^{166}$Yb and
$^{164,166}$Er nuclei at $\hbar \omega_x \approx 0.53$~MeV [see Figs.\
\ref{fig:CRHB-Tm}(a), (e) and (f)]; these nuclei are the
neighbors of $^{168}$Yb. These frequencies are only slightly lower than the
one at which experimental second upbending is seen in $^{168}$Yb.
In addition, second upbendings are predicted in $^{172}$Yb
and $^{170}$Er [see Figs.~\ref{fig:CRHB-Tm}(d) and (h)]. In the CRHB+LN
calculations, both sharp and gradual second upbendings appear in this
mass region contrary to only sharp first upbendings.

The differences (especially those related to different crossing
frequencies) in the model predictions obtained with the  NL1 and
NL5(E) CEDFs are attributable to the differences in the underlying
single-particle structure and, in particular, to the energies with
respect of vacuum of aligning orbitals which are responsible for
band crossing (see Fig.\ \ref{fig:routhians} below).
As illustrated by orange dashed
line in Fig.\ \ref{fig:CRHB-Tm}(a), some additional improvement in the description
of experimental data could be obtained by further optimization of
pairing. In the `NL5(E) (opt. pairing)' calculations, the scaling factor
$f$ of the Gogny pairing is selected in such a way that it reproduces
exactly the experimental MOI at $\hbar \omega_x =0.05$~MeV.  This
leads to both much better description of absolute values of MOI before
and after first band crossing in $^{166}$Yb and the frequency of first
band crossing.

All these results demonstrate that the LN method is a reasonably good approximation
to exact particle-number conserving method at least for the yrast bands in even-even nuclei.
It allows to avoid the pairing collapse (appearing in the standard BCS or HFB approaches)
in significant frequency range.
This collapse in the CRHB+LN calculations takes place only at very high rotational frequencies
where the pairing is very weak.
Note that at these frequencies, the calculations without pairing represent
a feasible alternative for the analysis of rotational properties.
In addition, at these frequencies such calculations are to a large degree free from
numerical or convergence problems existing both in the CRHB+LN and the cranking CDFT-SLAP approaches.

\begin{figure*}[!]
\includegraphics[width=0.95\textwidth]{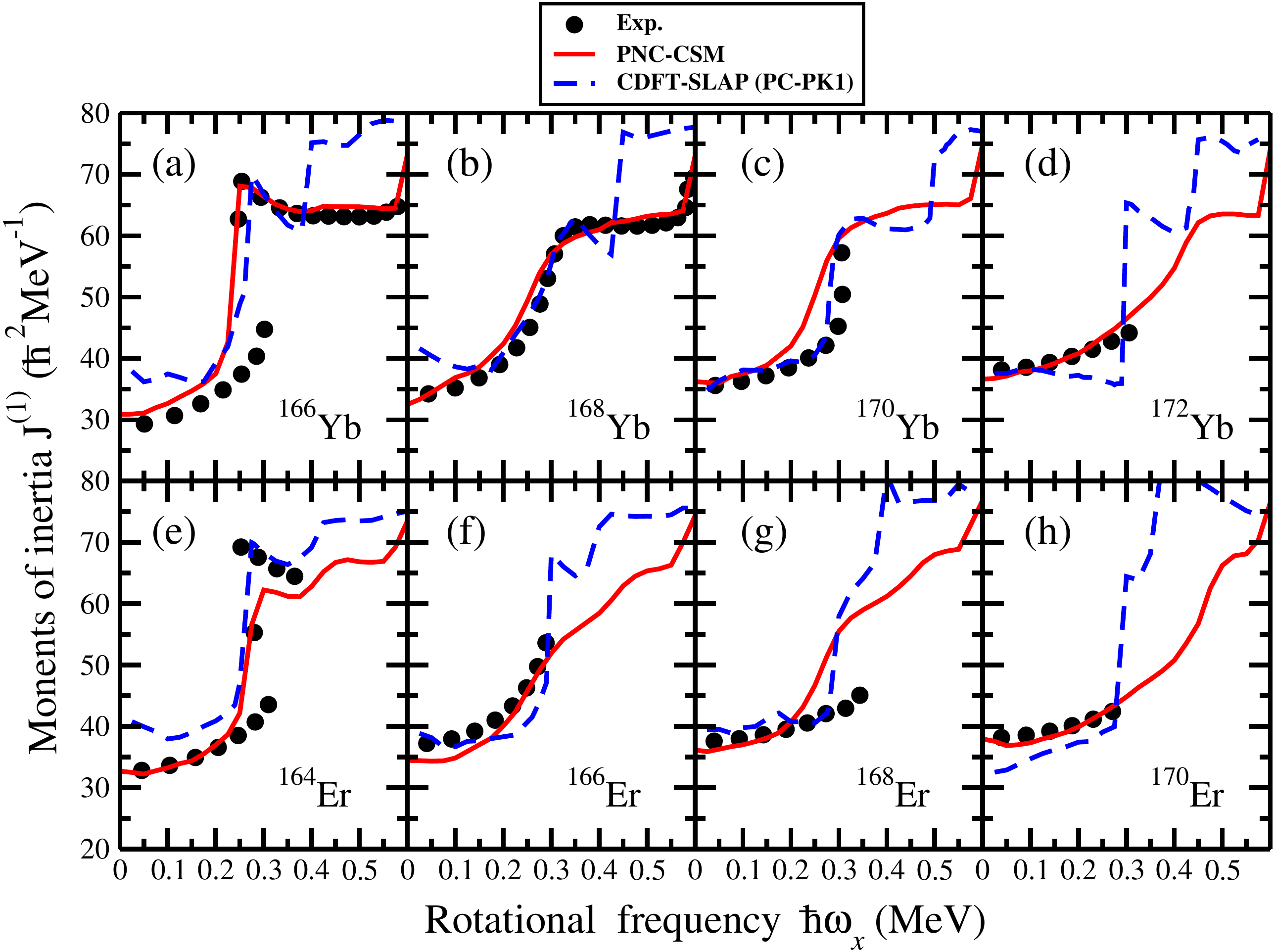}
\caption{\label{fig:pcpk1}
The same as in Fig.~\ref{fig:CRHB-Tm} but for the results of the
calculations obtained  with cranking CDFT-SLAP (blue dashed lines)
and PNC-CSM (red solid lines).
The PC-PK1 functional~\cite{Zhao2010_PRC82-054319} is used in the cranking CDFT-SLAP
calculations.
}
\end{figure*}

For a detailed comparison of the description of rotational properties
by different models,  Fig.~\ref{fig:pcpk1} presents the results of the
calculations obtained by the cranking CDFT-SLAP and PNC-CSM
for the same set of nuclei as shown in Fig.~\ref{fig:CRHB-Tm}.
These two models with exact particle-number conservation can
reproduce the experimental data reasonably well.
On average, the PNC-CSM reproduces the experimental MOIs better than the
cranking CDFT-SLAP. On the contrary, the accuracy of the
description of available experimental data by PNC-CSM and CRHB+LN
models is on average comparable (compare Figs.~\ref{fig:pcpk1} and~\ref{fig:CRHB-Tm}).
However, in general the predictions of these two models differ above
$\hbar \omega_x > 0.35$ MeV, especially in the Er isotopes.

The cranking CDFT-SLAP and CRHB+LN calculations share the common
feature: most of first band crossings are sharp and take place around
$\hbar \omega_x \sim 0.3$~MeV (see Figs. \ref{fig:CRHB-Tm} and \ref{fig:pcpk1}).
On the contrary, in the PNC-CSM calculations the upbendings are sharp only
for the $N=96$ isotones and they become more gradual with increasing neutron number.
As mentioned in Refs.~\cite{Wu1991_PRL66-1022, Wu1991_PRC44-2566},
the yrast-yrare interaction strength, responsible for band crossing
features, depends sensitively on the occupation number distribution in the
high-$j$ orbitals.
As a result,  the differences in band crossing features
may come from the differences in the single-particle structure
obtained  by different models and the rate of their change
in the band crossing region.
For example, the deformations (and thus the mean field) are fixed in the PNC-CSM calculations.
Thus, the single-particle structure changes gradually at the band crossing
region and the upbendings tend to be more gradual.
On the contrary, the mean field is defined fully self-consistently with
rotational frequency in the CDFT-based calculations.
As a consequence, the upbendings may lead to a substantial
change of equilibrium deformation and thus to significant changes of
the single-particle structure (see the discussion of the
Figs.~\ref{fig:bg} and \ref{fig:routhians} below).
Therefore, the interaction between the the
GSB and $s$-band configurations in the band crossing
region may be weak, which will lead to a sharp upbending
in the cranking CDFT-SLAP and CRHB+LN calculations.

It is necessary to recognize that spectroscopic quality of CEDFs is
lower than that of phenomenological potentials such as
Nilsson potential~\cite{Afanasjev2011_PLB706-177, Dobaczewski2015_NPA944-388}.
This is because CEDFs are fitted only to bulk properties (such as nuclear
masses and charge radii in the case of the PC-PK1 functional) and no information
on single-particle energies is used in the fitting protocols.
On the contrary, the set of Nilsson parameters used in the present manuscript
is fitted to the energies of the 1-qp states in the mass region under study.
These facts may also contribute into the differences, related to the first
band crossing features, existing between CDFT-based and Nilsson potential based models.
Moreover, the differences in the type of employed pairing force (Gogny
pairing in CRHB+LN versus monopole pairing in cranking CDFT-SLAP and
versus monopole+quadrupole pairing in PNC-CSM) and the way
particle number projection is treated also can play a role in above discussed
differences between model predictions.

\begin{figure}[!]
\centering
\includegraphics[width=0.95\columnwidth]{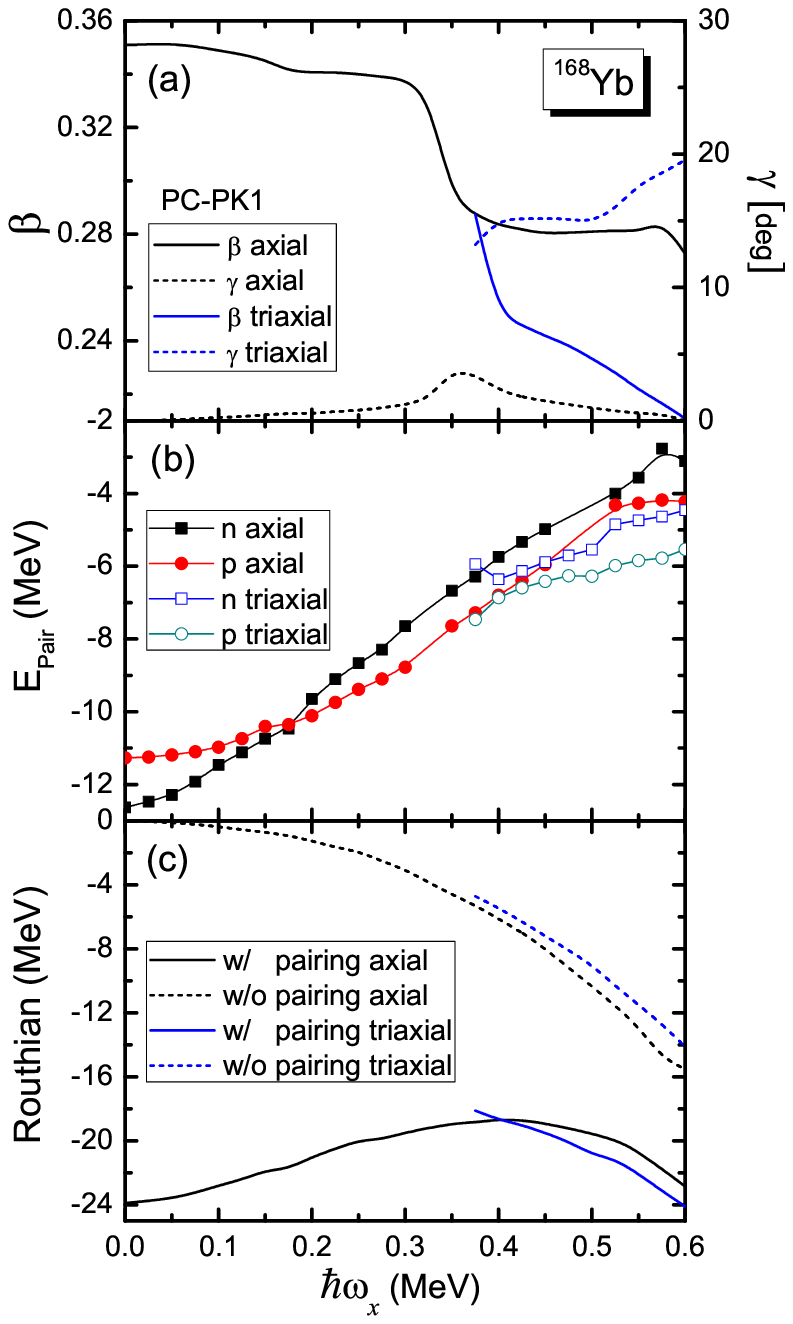}
\caption{\label{fig:168yb}
The evolution of the deformation parameters
($\beta,\gamma$) [panel (a)],
proton and neutron pairing energies [panel (b)],
and total Routhians [panel (c)]
with rotational frequency obtained in the cranking CDFT-SLAP
calculations with CEDF PC-PK1 for two competing at high spin
minima in $^{168}$Yb.
One minimum [denoted as `axial'] has axial or near-axial shapes,
while another [denoted as `triaxial'] corresponds to triaxial shapes with
 $\gamma \approx 18^{\circ}$.
The results of the calculations with and
without pairing are denoted as `w/' and `w/o', respectively.
Note that the same constant energy is subtracted in
all total Routhians, which sets the
total Routhian of  the minimum with near-axial shape
in the calculations without pairing to zero at
$\hbar\omega_x=0$.}
\end{figure}

The PNC-CSM calculations predict the existence of a sharp second
upbending in all Yb and Er isotopes at rotational frequency
$\hbar\omega_x~\sim0.58$~MeV (see Fig.~\ref{fig:pcpk1}).
For the case of $^{168}$Yb, this is consistent with available experimental data [see
Fig.~\ref{fig:pcpk1}(b)]. In the cranking CDFT-SLAP calculations, the second
upbending takes place at substantially lower frequencies as compared
with the PNC-CSM and CRHB+LN results.
This is due to the appearance of triaxial minimum with
$\gamma \approx 18^{\circ}$
at high rotational frequencies in the cranking CDFT-SLAP
calculations which competes with near-axial minimum.
Fig.~\ref{fig:168yb} shows the evolution of deformation parameters ($\beta,\gamma$),
proton and neutron pairing energies, and total Routhians with rotational frequency
obtained in the cranking CDFT-SLAP calculations
with PC-PK1 for these two minima in $^{168}$Yb.
The ground state of $^{168}$Yb is axially deformed.
With increasing rotational frequency, the triaxial deformation $\gamma$
gradually increases but still remains relatively small.
A triaxial minimum with the energy comparable to the
one of near-axial minimum develops
at $\hbar\omega_x \approx 0.35$~MeV
after the first band crossing.
It can be seen in Fig.~\ref{fig:168yb}(a) that this minimum has
substantially smaller quadrupole deformation $\beta$ than the
near-axial one and that the triaxial deformation increases from
$\gamma \approx 15^\circ$ at  $\hbar\omega_x \approx 0.35$~MeV
to  $\gamma \approx 20^\circ$ at  $\hbar\omega_x = 0.60$~MeV.
Fig.~\ref{fig:168yb}(c) shows the total Routhians of calculated configurations.
One can see that the total Routhian of near-axial
minimum is energetically favoured as compared with the one of
triaxial minimum in the calculations without pairing.
However, in the calculations with pairing the triaxial minimum
becomes energetically favoured at $\hbar\omega_x > 0.4$~MeV
because pairing energies in this minimum are substantially
larger than those in near-axial one [see Fig.~\ref{fig:168yb}(b)].
The energies of these two minima are very close to each other
in some rotational frequency range. Thus, the
self-consistent calculations should be carefully carried out to ensure
that the real global minimum is found.
Note that the competition of these two minima depends not only on the
details of the pairing interaction, but also on underlying single-particle
structure.

The calculated MOIs obtained in the cranking CDFT-SLAP are less smooth
as compared with the CRHB+LN ones. This is because in the cranking CDFT-SLAP,
the many-body Hamiltonian is diagonalized directly in the MPC space.
As a consequence, the eigenstate [Eq.~(\ref{eq:psi})] is no longer a Slater determinant but the
superposition of many Slater determinants. When investigating heavy nuclei with high
single-particle level densities, there may exist several low-lying MPCs with
very close excitation energies, especially when triaxial deformation appears.
With different initial mean field, the near degeneracy of these MPCs may lead
the cranking CDFT-SLAP calculations to converge to somewhat different
minima, which have slightly different expansion coefficients $C_i$ in the eigenstate [Eq.~(\ref{eq:psi})].
As a consequence, the change of rotational frequency can trigger minor
discontinuities in the occupation probabilities of the single-particle levels
located in the vicinity of the Fermi level. If some of these affected states
are high-$j$ ones, this can lead to small fluctuations in MOIs calculated
as a function of rotational frequency.
This defect of cranking CDFT-SLAP can be avoided by using the single-particle level
tracking technique and considering the
overlap between two eigenstates  calculated at
adjacent rotational frequencies~\cite{Meng2006_PRC73-037303, Shi2018_PRC97-034317}.
However, it is too time-consuming for a systematic investigation of these heavy nuclei.

\begin{figure*}[!]
\centering
\includegraphics[width=0.95\textwidth]{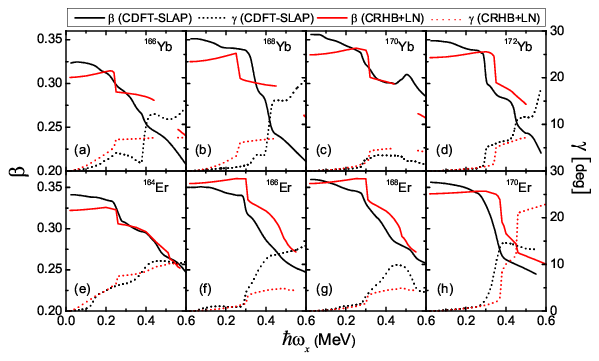}
\caption{\label{fig:bg}
The evolution of deformation parameters $\beta$ and $\gamma$
with rotational frequency in even-even nuclei under study obtained in
the cranking CDFT-SLAP  (solid lines) and CRHB+LN (dotted lines)
calculations. The cranking CDFT-SLAP and CRHB+LN calculations are
performed with the PC-PK1 and NL5(E) CEDFs, respectively.
}
\end{figure*}

Figure~\ref{fig:bg} shows the evolution of deformation parameters $\beta$
and $\gamma$ with rotational frequency obtained in two CDFT-based approaches.
One can see that in general it is similar in both approaches. However, some
differences are also present.
At low frequencies, all nuclei are axially symmetric and
with exception of $^{166}$Er quadrupole deformations $\beta$
obtained in cranking CDFT-SLAP calculations are somewhat larger
than those calculated in the CRHB+LN approach.
The triaxiality gradually increases with increasing rotational frequency up to
first band crossing in both calculations.  However, in this frequency
range the behaviour of calculated quadrupole deformations is
different in the cranking CDFT-SLAP and CRHB+LN approaches. With
increasing rotational frequency up to first band crossing, the $\beta$
values gradually decrease/increase in the cranking CDFT-SLAP/CRHB+LN
calculations.  Similar trend of the evolution of the $\beta$ values
with increasing rotational frequency has also been seen in other
CRHB+LN calculations~\cite{Afanasjev1999_PRC60-051303} and
in non-relativistic cranked HFB calculations~\cite{Terasaki1995_NPA593-1}.
Both calculations show that in the first
band crossing region of these
nuclei the quadrupole deformations $\beta$ rapidly decrease and triaxial
deformations $\gamma$ quickly increase.
As a result of these significant deformation changes, the first
band crossing is calculated in these two CDFT-based approaches
to be sharp  in most of the cases. The second band crossing leads to a further
decrease of quadrupole deformation. With a pair of exception, it also triggers
further increase of $\gamma$-deformation.
Fig.~\ref{fig:bg} shows that both CDFT-based approaches predict significant
triaxial deformation in these nuclei after the first band crossing.
However, due to non-selfconsistent nature of the cranked shell model,
the deformation is an input parameter in the PNC-CSM and the model does
not allow the variation of deformation with spin.
Thus, the axial symmetry is assumed in PNC-CSM calculations
and the magnitude of the quadrupole deformation is taken from
microscopic+macroscopic calculations which have similar structure of the single-particle potential.
This is also consistent with experimental information on axial symmetry
of the ground states in the rare-earth nuclei under study
as well as with the results of two CDFT-based model calculations for the ground states.
Note that the axial symmetry is adopted in absolute majority of cranked shell model
calculations for the rare-earth nuclei under study (see, for example, Ref.~\cite{Asgar2017_PRC95-031304R}).

Some differences seen in the results of the cranking  CDFT-SLAP and CRHB+LN
calculations emerge from different employed CEDFs. For example, the
$\gamma$-deformations of the solutions obtained after second band
crossing are typically larger in the cranking CDFT-SLAP calculations. The pairing
is weak in this frequency range and thus these differences could not be
related to the treatment of pairing or the selection of the pairing force.

\begin{figure}[h]
\includegraphics[width=0.95\columnwidth]{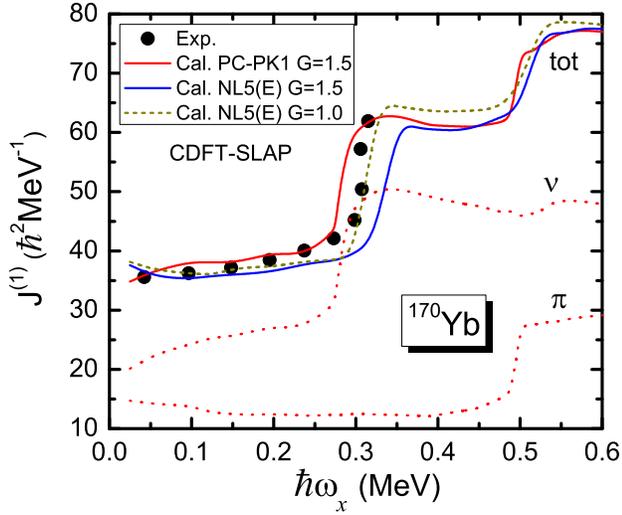}
\caption{\label{fig:170yb-moi}
The MOIs of the GSB in $^{170}$Yb obtained
in the cranking CDFT-SLAP approach with the PC-PK1 and NL5(E) CEDFs.
The contributions from proton and neutron subsystems
to the MOIs obtained by PC-PK1 CEDF
are shown by red dotted lines.
}
\end{figure}

\begin{figure}[h]
\includegraphics[width=1.0\columnwidth]{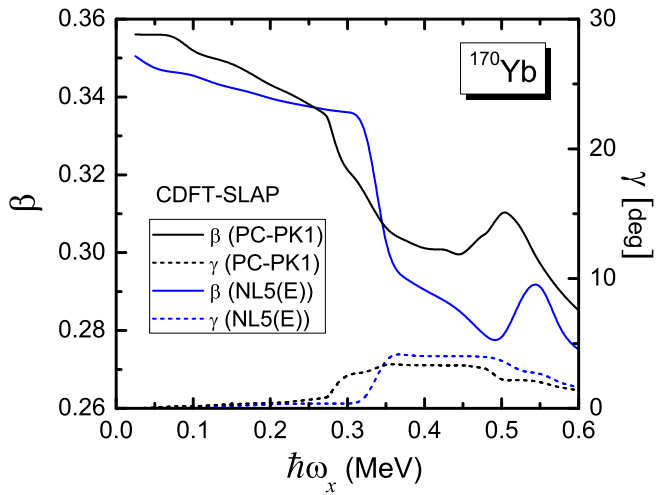}
\caption{\label{fig:170yb-def}
The evolution of deformation parameters $\beta$ and $\gamma$
with rotational frequency in $^{170}$Yb obtained
in the cranking CDFT-SLAP calculations with the NL5(E) and PC-PK1 CEDFs.
}
\end{figure}

 Figure~\ref{fig:170yb-moi}  compares the results of cranking CDFT-SLAP
calculations for the GSB in $^{170}$Yb obtained with PC-PK1 and NL5(E)
functionals. For the same pairing strength $G_{\rm n}=G_{\rm p}=1.5$~MeV,
the MOIs obtained before first upbending with NL5(E) are somewhat smaller than those obtained
with PC-PK1 and experimental ones. The calculations with NL5(E)/PC-PK1
slightly underestimate/overestimate the first band crossing frequency.
Since the single-particle level structures located in the vicinity of the Fermi surface
are different in those two CEDFs, the corresponding pairing strengths should
not necessary be the same.
The reduction of proton and neutron pairing
strengths to $G_{\rm n}=G_{\rm p}=1.0$~MeV in the calculations with NL5(E) leads to a visible
improvement of the description of experimental data (see Fig.~\ref{fig:170yb-moi}).
Note that the equilibrium deformations obtained in the calculations with PC-PK1 and NL5(E)
are rather close to each other (see Fig.~\ref{fig:170yb-def}).
It also can be seen that
the first upbending is caused by the contribution from neutron subsystem,
and the second upbending is caused by the contribution from proton subsystem.
The same conclusion can be obtained for all even-even Er and Yb isotopes
investigated in the present work by cranking CDFT-SLAP.

With the exception of first band crossing region, the behavior of
the calculated MOIs presented in Fig.~\ref{fig:170yb-moi} are very close to each other.
This is a consequence of the fact that the rotation is a collective phenomenon
built on the contributions of many single-particle orbitals.
As a result, minor differences in the single-particle structure introduced
by the use of different functionals do not lead to substantial changes in MOIs.
The only exception is the band crossing region which is defined
by the alignment of selected pair of the orbitals and which depends more on
the energies and alignment properties of this pair.
Note that these features are also observed in the CRHB+LN calculations
(see Fig.~\ref{fig:CRHB-Tm}).

\begin{figure*}[ht]
\includegraphics[width=0.95\columnwidth]{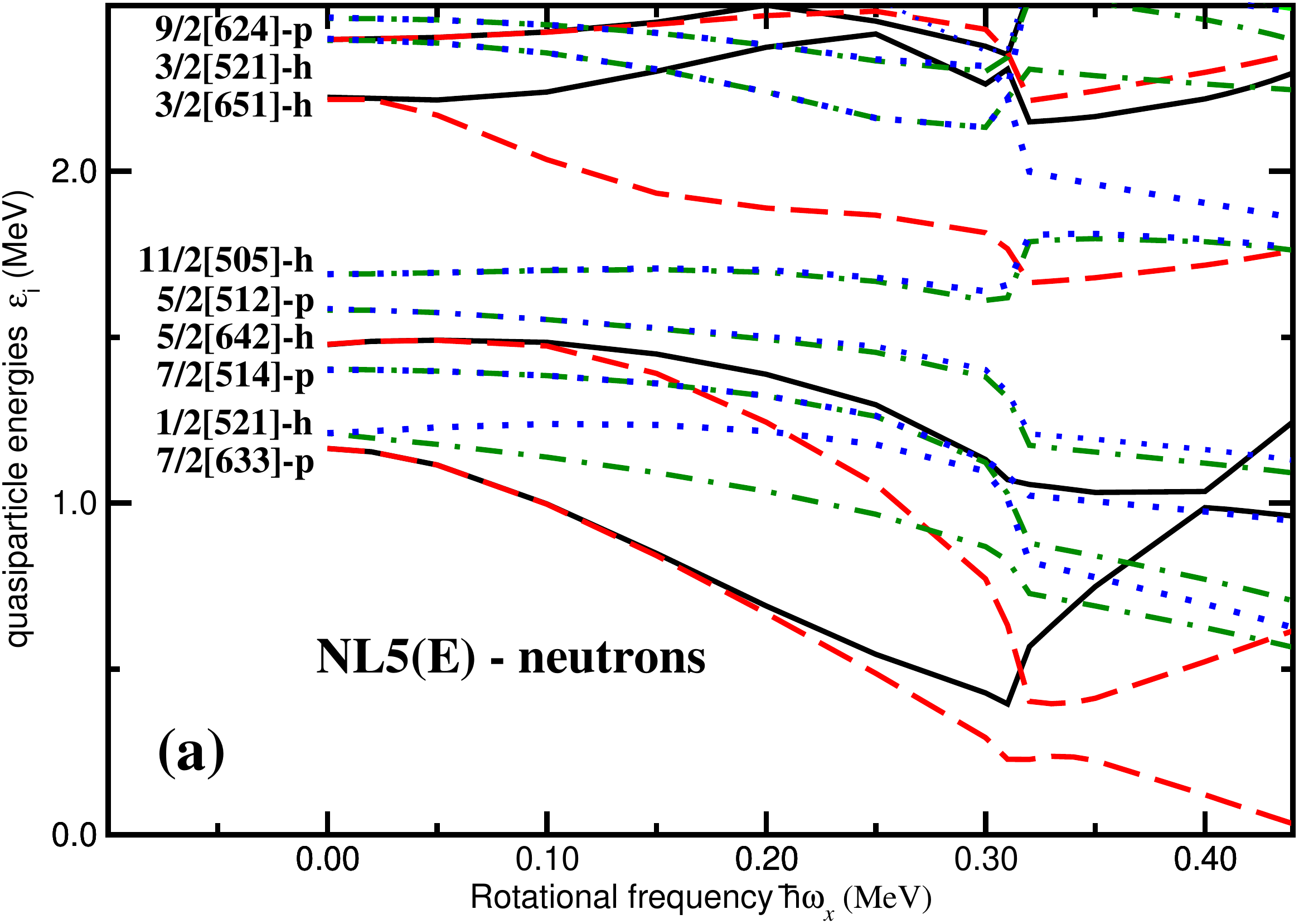}
\includegraphics[width=0.95\columnwidth]{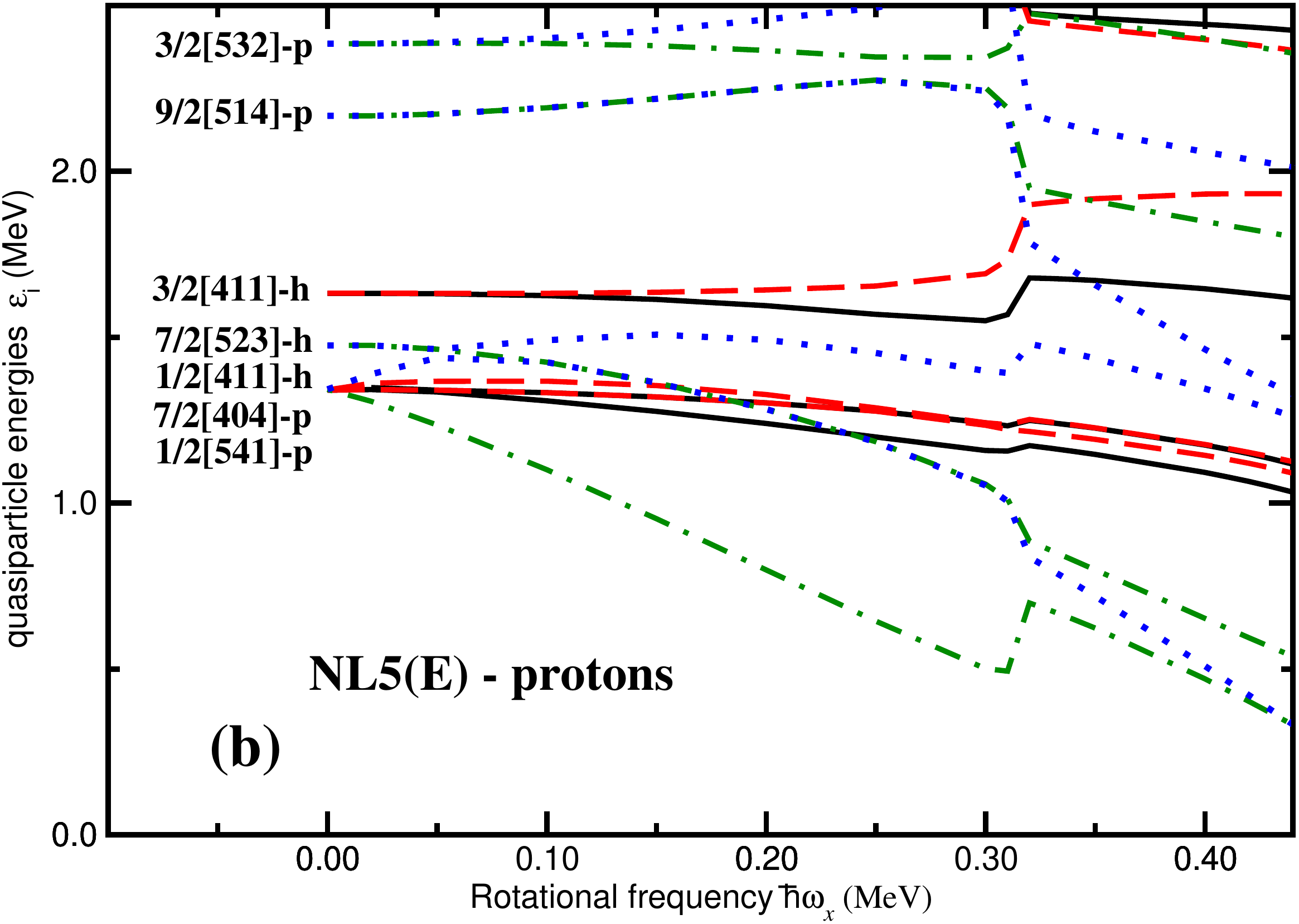}
\includegraphics[width=0.95\columnwidth]{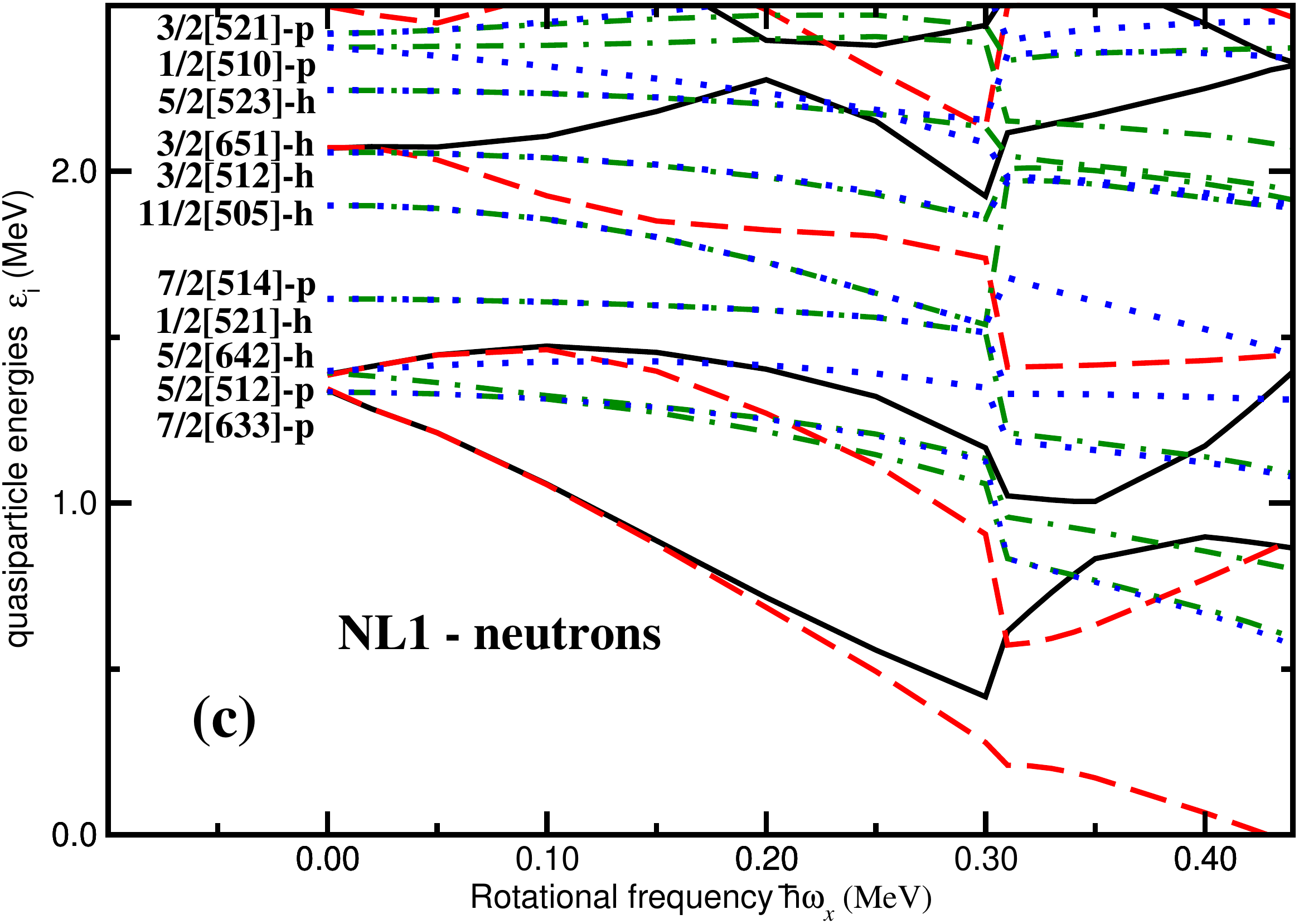}
\includegraphics[width=0.95\columnwidth]{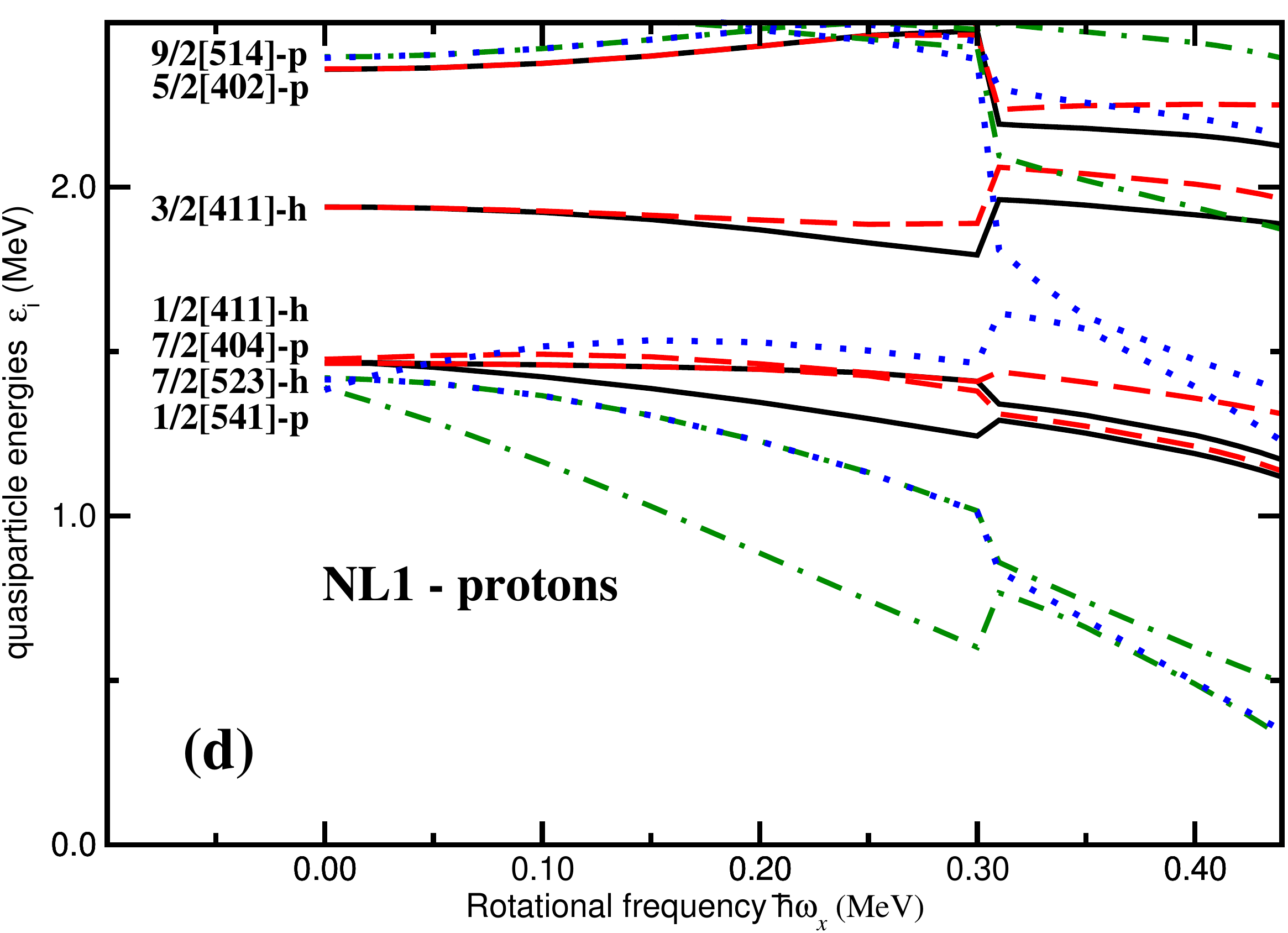}
\caption{\label{fig:routhians}
Neutron and proton quasiparticle energies (routhians) corresponding
to the lowest configurations in $^{170}$Yb obtained in the CRHB+LN calculations
with  NL5(E) (top panels) and NL1 (bottom panels) CEDFs. They are given along the
deformation path of these configurations.
Long-dashed, solid, dot-dashed and dotted lines indicate  $(\pi=+, r=+i)$,
$(\pi=+, r=-i)$, $(\pi=-, r=+i)$ and $(\pi=-, r=-i)$ orbitals, respectively.
At $\Omega_x=0.0$ MeV, the quasiparticle orbitals are labeled by the asymptotic
quantum numbers $[Nn_z\Lambda]\Omega$ (Nilsson quantum numbers) of the
dominant component of the wave function. The letters `p' and `h' before the
Nilsson labels are used to indicate whether a given routhian is of particle
($V^2 <0.5$)
or hole ($V^2 > 0.5$) type.
}
\end{figure*}

There is a substantial difference between the CRHB+LN and cranking CDFT-SLAP
calculations in respect  of the modifications of the calculated MOIs
induced by the change of pairing strength.  Fig.~\ref{fig:170yb-moi} shows that in the cranking CDFT-SLAP
calculations with NL5(E) CEDF the reduction of monopole pairing strength by
1/3 leads only to moderate change in the MOI.
Similar features have been observed in the cranking CDFT-SLAP calculations in other mass
regions~\cite{Liu2019_PRC99-024317}.
On the contrary, the increase of scaling factor $f$ of the Gogny pairing force from 0.950
[blue dashed line in Fig.~\ref{fig:CRHB-Tm}(a)] to
0.998 [orange dashed line in Fig.~\ref{fig:CRHB-Tm}(a)] leads to
larger changes in calculated MOI.
In a similar fashion, the 10\% change in pairing  strength of the Gogny pairing force
leads to a substantial changes in the calculated MOIs of superdeformed bands of the
$A\sim 190$ mass region (see Fig.~12 in Ref.~\cite{Afanasjev2000_NPA676-196}).

Figure~\ref{fig:routhians} shows the quasiparticle routhians
obtained in the CRHB+LN calculations with the NL1 and NL5(E) functionals.
Although the energies of the routhians
with the same structure are somewhat different in these functionals,
there are large similarities in the general structure of the quasiparticle spectra
obtained with these two functionals. For example, the alignments of the
quasiparticle orbitals, reflected in the energy slope of their
routhians as a function
of rotational frequency, are very similar in both functionals. In addition, a similar sets
of proton and neutron quasiparticle states  appear in the vicinity of the
Fermi level in these CEDFs.  Moreover, in both functionals, the first paired band
crossing is due to the  alignment of the neutron $7/2^+[633]$ orbitals.

\begin{figure*}[!]
\centering
\includegraphics[width=0.95\textwidth]{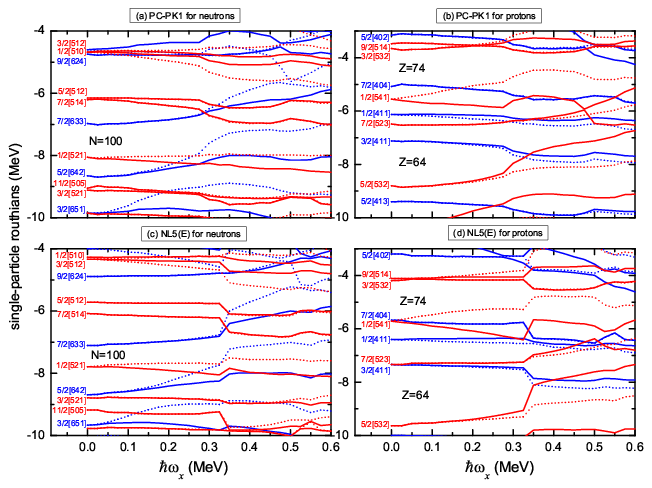}
\caption{\label{fig:slap-spl}
Neutron (left panels) and proton (right panels) single-particle routhians
located in the vicinity of the Fermi level of the $^{170}$Yb as a function of rotational frequency.
Positive (negative) parity routhians are shown by blue (red) lines.
Solid (dotted) lines are used for signature $\alpha=+1/2$ ($\alpha=-1/2$).
}
\end{figure*}

Figure~\ref{fig:slap-spl} shows the single-particle routhians obtained in the cranking
CDFT-SLAP calculations with PC-PK1 (upper panels) and NL5(E) (lower panels) functionals.
There are large similarities between these two functionals in terms of the
locations of similar set of the single-particle states in the vicinity of the Fermi level,
the signature splittings of single-particle orbitals and their evolution with rotational
frequency and the slope of the single-particle energies with rotational frequency.
The comparison of the quasiparticle routhians shown in Figs.~\ref{fig:routhians}(a)
and (b) and the single-particle routhians displayed in Figs.~\ref{fig:slap-spl}(b) and
(d) allows to establish close correspondence between underlying single-particle
structure obtained in the CRHB+LN and cranking CDFT-SLAP calculations with the NL5(E)
functional. First band crossing leads to sharper changes in the energies of
the proton and neutron single-particle states in the cranking CDFT-SLAP calculations with
NL5(E) as compared with those for PC-PK1 (see Fig.~\ref{fig:slap-spl}).
This is due to more drastic deformation changes obtained in the band crossing region
in the calculations with NL5(E) (see Fig.~\ref{fig:170yb-def}).
Note also that first band crossing takes place
at higher frequency in NL5(E) than in PC-PK1.

\begin{figure}[!]
\includegraphics[width=0.95\columnwidth]{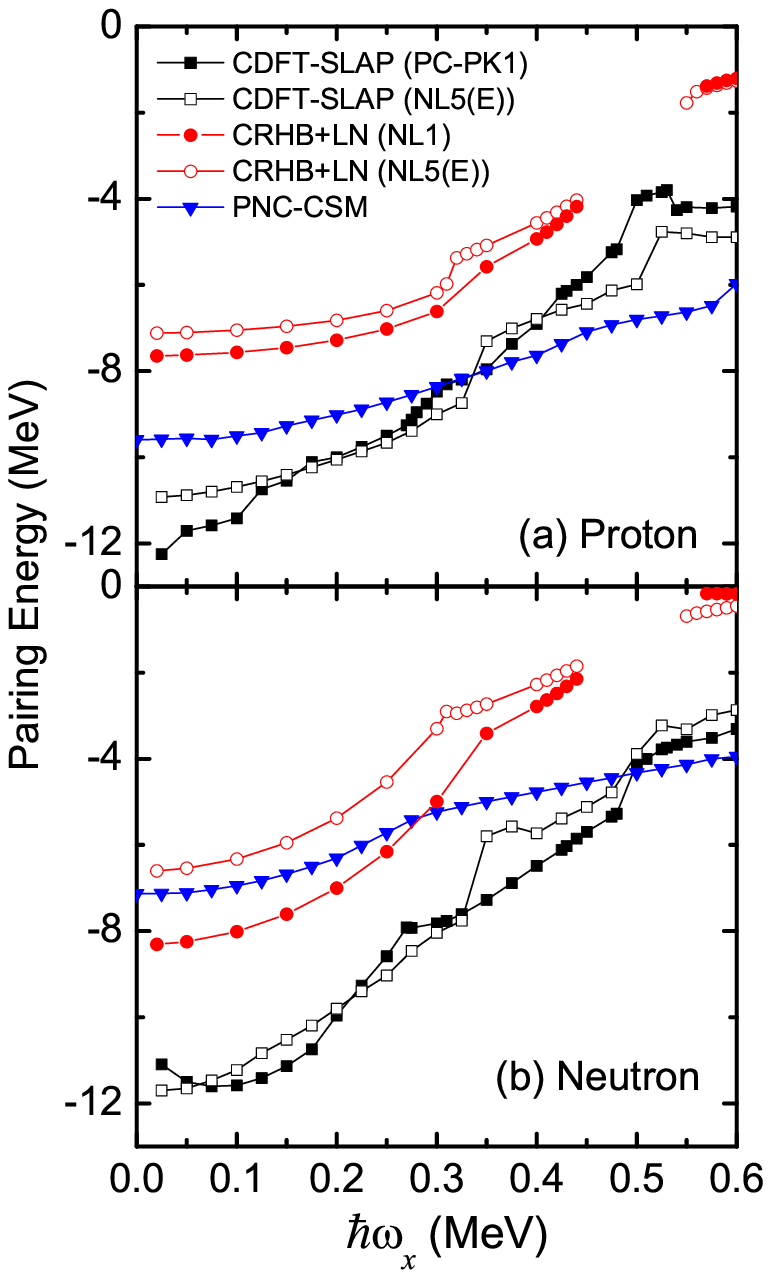}
\caption{\label{fig:170yb-pairing}
The neutron and proton pairing energies obtained in different calculations as a
 function of  rotational frequency in $^{170}$Yb.
}
\end{figure}

Figure~\ref{fig:170yb-pairing} shows the pairing energies for neutron
and proton subsystems of $^{170}$Yb as a function of rotational
frequency obtained in the CRHB+LN, cranking CDFT-SLAP and PNC-CSM calculations.
In general, both proton and neutron paring energies decrease with
rotational frequency but even in the high-spin region they are still non-zero.
Paired band crossings trigger some reduction of pairing
energies and the change of the slope of pairing energies as a function
of rotational frequency  in the CDFT-based approaches.
This  is due to the change of the deformation of the mean-fields
(see Fig.~\ref{fig:bg}), the quasiparticle energies (see Fig.\ \ref{fig:routhians})
in the CRHB+LN approach and the single-particle Routhians (see Fig. \ref{fig:slap-spl})
in the cranking CDFT-SLAP approach taking place at the band crossings.
Although the methods of the treatment of pairing correlations are exactly
the same in the cranking CDFT-SLAP and PNC-CSM approaches,
the variations of calculated pairing energies with rotational frequency are
different. Contrary to the cranking CDFT-SLAP approach,  the pairing energies
decrease smoothly (even in the band crossing regions) with increasing
rotational frequency in the PNC-CSM calculations.
This is because the deformation is fixed in the PNC-CSM calculations.
As a consequence, the single-particle levels
change gradually with rotational frequency.

The calculated pairing energies depend both
on theoretical framework as well as on the employed functional.
The former dependence is due to different definitions of pairing energies
in the CRHB+LN and cranking CDFT-SLAP approaches [compare Eqs.~(\ref{Epair})
and (\ref{Epair-CDFT-SLAP})] and the use of different pairing forces.
The latter one enters through the dependence of pairing energies
on the single-particle level density: in extreme case of large shell
gap in the vicinity of the Fermi level there will be pairing collapse
(see Ref.\ \cite{Shimizu1989_RMP61-131}).
For example, the CRHB+LN calculations with the NL1 and NL5(E) CEDFs are
performed with comparable scaling factors $f$ of the Gogny pairing force.
As a consequence, the similarity/difference of proton/neutron
pairing energies in these calculations (see Fig.\ \ref{fig:170yb-pairing})
are due to similarity/difference of the density of the proton/neutron
single-particle states in the vicinity of respective Fermi levels (see
Fig.\ \ref{fig:routhians}). The situation is the same for the
cranking CDFT-SLAP calculations with the PC-PK1 and NL5(E) CEDFs.

\section{Rotational properties of odd-proton nuclei $^{165,167,169,171}$Tm}
\label{Sec:PNC}

The rotational structures in odd-mass nuclei provide additional
testing ground for theoretical approaches.
In addition, they yield important information on underlying single-particle structure,
thus providing an extra tool for the configuration
assignment (see discussion in Sec.~4C of Ref.~\cite{Afanasjev2013_PRC88-014320}).
However, the calculations in the cranking CDFT-SLAP and CRHB+LN approaches in such
nuclei are extremely time consuming requiring significantly larger computational time than
similar calculations in even-even nuclei.

\begin{figure*}[!]
\includegraphics[width=0.9\textwidth]{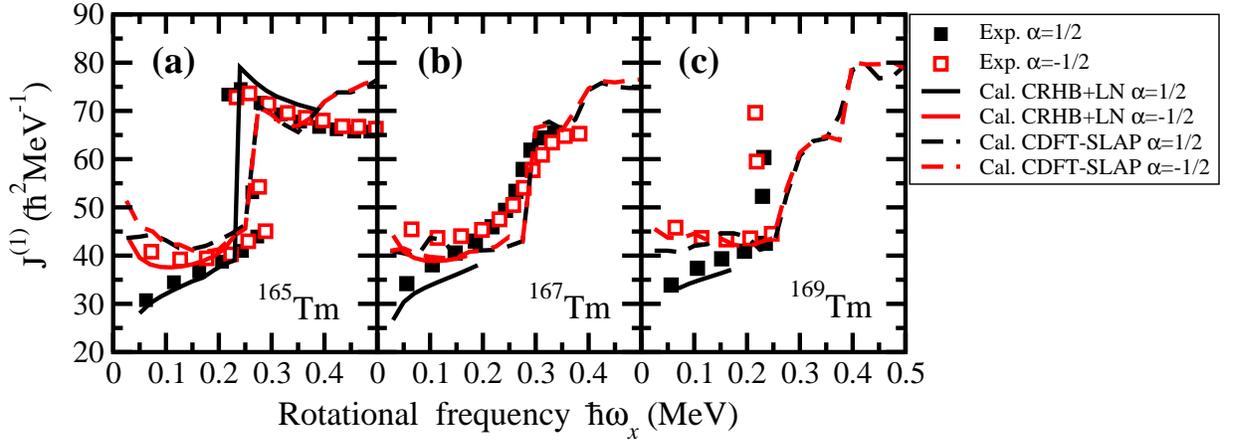}
\caption{\label{fig:odd}
The MOIs of the GSB $\pi 1/2^+[411]$ in odd-$A$ Tm isotopes.
Experimental data, taken from Refs.~\cite{Jensen2001_NPA695-3,
Burns2005_JPG31-S1827, Asgar2017_PRC95-031304R} are compared with
the results of the CRHB+LN with NL5(E) and the
cranking CDFT-SLAP calculations with PC-PK1.
}
\end{figure*}

In addition, there is a principal difference between the calculations of odd-mass
nuclei in the CRHB+LN and the cranking CDFT-SLAP approaches.
Such calculations in the CRHB+LN approach (as well as in non-relativistic HFB based approaches)
employ blocking of specific single-particle orbital(s) for definition of nucleonic configurations.
However, this frequently leads to numerical instabilities emerging from the interaction of
blocked orbital with other single-particle orbital having the same quantum numbers
and located close in energy (see Ref.~\cite{Afanasjev2013_PRC88-014320}).
This deficiency is clearly seen in Fig.~\ref{fig:odd} where numerical convergence has been
obtained in restricted frequency range for the $\pi 1/2^+[411]$ GSBs of odd-$A$ Tm
isotopes and mostly for the $\alpha=+1/2$ signature. Note that calculated results are
reasonably close to experimental data.
Such numerical instabilities are also a reason why the calculations of rotational structures
in odd-$A$ and odd-odd nuclei in relativistic and non-relativistic density functional theories
are very rare. To our knowledge, such calculations have been performed so far only for few
such nuclei (mostly for actinides) in Refs.~\cite{Bender2003_NPA723-354,
OLeary2003_PRCC67-021301, Herzberg2009_EPJA42-333, Jeppesen2009_PRC80-034324,
Afanasjev2013_PRC88-014320} and mostly in the CRHB+LN framework.

On the contrary, the specific orbital is not blocked in shell-model based
approaches and the process for calculating odd-$A$ nuclei is exactly the
same as in even-even ones in the cranking CDFT-SLAP.
Thus, there is no numerical convergence problems typical for HFB approaches.
The analysis of the occupation probabilities of the
single-particle levels located in the vicinity of the Fermi level allows to define nucleonic
configurations.
However, the problems similar to those revealed in the discussion of
Fig.~\ref{fig:168yb} and emerging from the convergence of the calculations to slightly different
minima exist also in odd-$A$ nuclei. They increase computational time and require substantial time for
the analysis of the calculations and configuration assignment to observed band.

Figure~\ref{fig:odd} compares experimental data on MOIs
of the GSB $\pi 1/2^+[411]$ in odd-$A$ Tm isotopes with the results of the
calculations of the CDFT-based models.
In the cranking CDFT-SLAP calculations, the convergence can be obtained up to
very high frequency in all nuclei under study (see Fig.~\ref{fig:odd}). The frequency of first
band  crossing and the MOIs immediately after it are very close to experimental data in
$^{165}$Tm. However, at low frequency the MOIs are somewhat overestimated in the
calculations  and the signature splitting is not reproduced. The latter feature is due to small
signature splitting of the $\pi 1/2^+[411]$ orbital obtained in the cranking CDFT-SLAP
calculations (see Fig.~\ref{fig:slap-spl}).
In addition, the cranking CDFT-SLAP calculations predict a second upbending at
$\hbar\omega_x \sim 0.4$~MeV (similar to the one predicted in
even-even nuclei in Fig.~\ref{fig:pcpk1}),
which is not observed in experiment.
In $^{167,169}$Tm nuclei, the calculated results are similar to those
obtained in $^{165}$Tm.

The MOIs of opposite signatures of the $\pi 1/2^+[411]$
band in $^{165}$Tm are rather well
reproduced before band crossing  in the CRHB+LN calculations
with the NL5(E) functional (see Fig.~\ref{fig:odd}).
However, at higher frequency only the $\alpha=+1/2$ branch converges in
the CRHB+LN calculations and only for rotational frequencies $\hbar \omega_x < 0.38$~MeV.
For this branch, the calculated upbending takes place at the frequency which is close
to medium frequency of experimental backbending.
In $^{167}$Tm, the calculations converge only up to $\hbar \omega_x \sim 0.19$~MeV.
The signature splitting is rather well reproduced but the calculations somewhat
underestimate the experimental values of MOIs.
In $^{169}$Tm, the CRHB+LN calculations converge only for $\alpha=+1/2$ branch and only
for low frequencies. Here the results of the calculations are very close to experimental data.
Note that no convergence for the $\pi 1/2^+[411](\alpha=\pm 1/2)$ bands have
been obtained in the CRHB+LN calculations with the NL1 functional. The close energies of
the  $\pi 1/2^+[411]$ and $\pi 7/2^+[404]$ quasiparticle orbitals [see Figs.~\ref{fig:routhians}
(b) and (d)], leading to a substantial interaction between them, is the most likely source of
the convergence problems observed in the CRHB+LN calculations.

In the light of time-consuming nature of the calculations within
the CDFT-based approaches and above discussed technical difficulties,
the systematic investigation of the properties of
odd-proton nuclei $^{165,167,169,171}$Tm will be performed here
only in the PNC-CSM framework.

\begin{table*}[!]
\caption{\label{tab:qpTm}
The comparison between experimental and
PNC-CSM results for bandhead energies of low-lying 1- and 3-qp states
in $^{165,167,169,171}$Tm.
$E_{\rm th}$, $E_{\rm stand}$, and $E_{\rm A150}$ denote the
calculated energies obtained with different sets of the Nilsson parameters.
Here, `th', `stand' and `A150' stand for the parameters
adopted in the present work, derived for global description in
Ref.~\cite{Bengtsson1985_NPA436-14} (frequently called `standard
Nilsson parameters') and defined from high spin properties in the
$A\sim 150$ mass region in Ref.~\cite{Bengtsson1990_NPA512-124}, respectively.
The experimental data are taken from Refs.~\cite{Jain2006_NDS107-1075, Baglin2000_NDS90-431,
Baglin2008_NDS109-2033, Baglin2018_NDS151-334}.
$K^\pi=17/2^+$ and $K^\pi=17/2^-$ in $^{165}$Tm denote two 3-qp states
with the configuration $\pi7/2^+[404] \otimes \nu^2 5/2^-[523]5/2^+[642]$ and
$\pi7/2^-[523] \otimes \nu^2 5/2^-[523]5/2^+[642]$, respectively.
}
\begin{center}
\begin{spacing}{1.2}
\def\temptablewidth{17cm}
\begin{tabular*}{\temptablewidth}{@{\extracolsep{\fill}}lllllll}
  \hline
  \hline
  Nuclei     & Configuration    & $E_{\rm exp}$~(keV)   & $E_{\rm th}$~(keV)
                                & $E_{\rm stand}$~(keV) & $E_{\rm A150}$~(keV)\\
  \hline
  $^{165}$Tm &  $\pi1/2^+[411]$ & 0     & 0     & 0     & 0    \\
  $^{165}$Tm &  $\pi7/2^+[404]$ & 80    & 93    & 393   & 482  \\
  $^{165}$Tm &  $\pi7/2^-[523]$ & 160   & 138   & 325   & 12   \\
  $^{165}$Tm &  $\pi5/2^+[402]$ & 315   & 414   & 722   & 465  \\
  $^{165}$Tm &  $\pi3/2^+[411]$ & 416   & 1312  & 1168  & 1488 \\
  $^{165}$Tm &  $\pi9/2^-[514]$ & 831   & 791   & 899   & 1272 \\
  $^{165}$Tm &  $17/2^+$
                                & 1634  & 1676  & 2020  & 2065 \\
  $^{165}$Tm &  $17/2^-$
                                & 1741  & 1721  & 1952  & 1595 \\
  $^{167}$Tm &  $\pi1/2^+[411]$ & 0     & 0     & 0     & 0    \\
  $^{167}$Tm &  $\pi7/2^+[404]$ & 180   & 335   & 549   & 664  \\
  $^{167}$Tm &  $\pi7/2^-[523]$ & 293   & 224   & 319   & 45   \\
  $^{167}$Tm &  $\pi5/2^+[402]$ & 558   & 680   & 859   & 618  \\
  $^{167}$Tm &  $\pi3/2^+[411]$ & 471   & 1451  & 1193  & 1336 \\
  $^{167}$Tm &  $\pi9/2^-[514]$ & 928   & 873   & 899   & 1284 \\
  $^{169}$Tm &  $\pi1/2^+[411]$ & 0     & 0     & 0     & 0    \\
  $^{169}$Tm &  $\pi7/2^+[404]$ & 316   & 542   & 733   & 826  \\
  $^{169}$Tm &  $\pi7/2^-[523]$ & 379   & 393   & 450   & 30   \\
  $^{169}$Tm &  $\pi5/2^+[402]$ & 782   & 916   & 1078  & 782  \\
  $^{169}$Tm &  $\pi9/2^-[514]$ & 1152  & 968   & 976   & 1226 \\
  $^{171}$Tm &  $\pi1/2^+[411]$ & 0     & 0     & 0     & 0    \\
  $^{171}$Tm &  $\pi7/2^+[404]$ & 636   & 655   & 856   & 945  \\
  $^{171}$Tm &  $\pi7/2^-[523]$ & 425   & 465   & 500   & 172  \\
  $^{171}$Tm &  $\pi5/2^+[402]$ & 912   & 1062  & 1254  & 899  \\
  $^{171}$Tm &  $\pi3/2^+[411]$ & 676   & 1560  & 1245  & 1442 \\
  \hline
  \hline
  \end{tabular*}
  \end{spacing}
  \end{center}
\end{table*}

Table~\ref{tab:qpTm} shows the comparison between the experimental and
calculated bandhead energies of the 1- and 3-qp states in $^{165,167,169,171}$Tm.
Note that the bandhead energies of the  $\pi 1/2^-[541]$ states are not shown
in this table due to the following reasons. First,  the deformation of this state is
larger than that for other states because it has  strong deformation driving
effect~\cite{Nazarewicz1990_NPA512-61, Warburton1995_NPA591-323}.
Second, because of strong decoupling effect arising from Coriolis interaction
the $I=5/2\hbar $ state is located lower in energy in experiment
than the bandhead with spin $I=1/2\hbar$.

One can see that calculated energies obtained with
`stand' and `A150' sets of the Nilsson parameters
(see caption of Table~\ref{tab:qpTm} for details)
cannot reproduce experimental data well.
This is especially true for the excitation energies of the $\pi7/2^+[404]$ state,
which are calculated too high in energy as compared with experimental data.
In addition, the sequence of the $\pi7/2^+[404]$ and $\pi5/2^+[402]$ states
is reversed as compared with experiment when the `A150' set of the Nilsson
parameters is used.
Note also that all three employed sets of the Nilsson parameters overestimate
experimental excitation energies of the $\pi3/2^+[411]$ states in all considered Tm isotopes.
The two 3-qp states observed in $^{165}$Tm are reproduced very well
by the Nilsson parameter set `th' adopted in the present work. On the
contrary, the energies of these states calculated with `stand' and `A150'
sets of the parameters deviate from experiment by 200-300~keV.
These results indicate that in general adopted set of the Nilsson parameters
improves a description of experimental data as compared with that
obtained with `stand' and `A150' sets of the parameters and
provides a reasonably accurate single-particle structure.
This  is important for a detailed investigation of rotational properties
and band crossing features of the nuclei under study.

\begin{figure}[!]
\includegraphics[width=1.0\columnwidth]{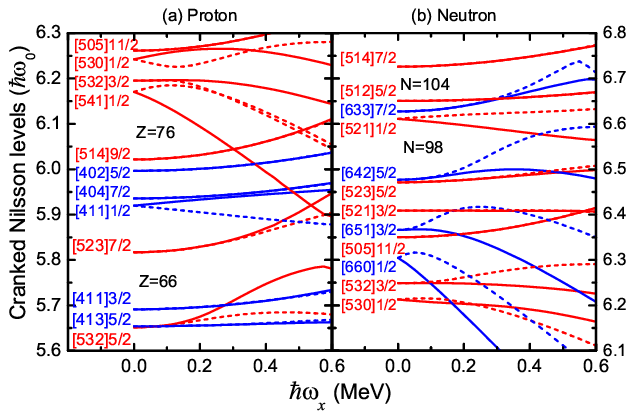}
\caption{\label{pnc-fig1:nil}
Proton [panel (a)] and neutron [panel (b)] single-particle routhians
located in the vicinity of the Fermi level of the $^{165}$Tm nucleus
as a function of rotational frequency $\hbar\omega_x$.
Positive (negative) parity routhians are shown by blue (red) lines.
Solid (dotted) lines are used for signature $\alpha=+1/2$ ($\alpha=-1/2$).
}
\end{figure}

Underlying single-particle structure and its evolution with rotational
frequency is exemplified in Fig.~\ref{pnc-fig1:nil}; similar structures are
also seen in the $^{167,169,171}$Tm nuclei.
At low rotational frequencies there exist a proton shell gap at $Z=76$
and two neutron shell gaps at $N=98$ and 104.
With increasing rotational frequency these gaps either disappear or get substantially reduced.
In the Tm isotopes of interest, with increasing neutron number
the neutron Fermi level is shifted up from $N=96$ to $N=102$.
Both the magnitude of the shell gaps and the position of the
Fermi level may affect the backbendings/upbendings in these Tm isotopes.
Note that the total Routhian surface (TRS) calculations of Ref.~\cite{Asgar2017_PRC95-031304R}
with Woods-Saxon potential show small neutron shell gap at
$N=102$ instead of the $N=104$ one present in our calculations.

\begin{figure*}[!]
\includegraphics[width=0.95\textwidth]{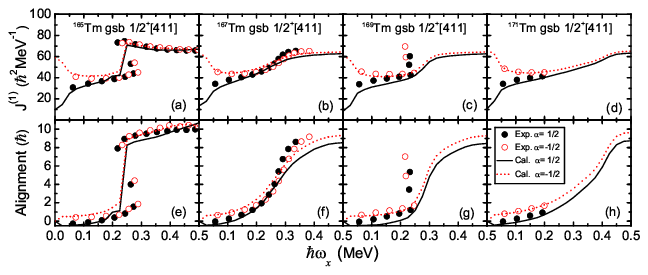}
\caption{\label{pnc-fig2:moi}
The experimental and calculated MOIs $J^{(1)}$ (upper panels) and alignments (lower panels)
of the GSBs $\pi 1/2^+[411]$  in $^{165,167,169,171}$Tm.
The experimental MOIs and alignments
are displayed
by black solid and red open circles for the $\alpha= +1/2$ and
$\alpha= -1/2$ branches of rotational band, respectively.
Corresponding calculated values are displayed by black solid
($\alpha= +1/2$) and red dotted ($\alpha= -1/2$) lines.
The experimental data are taken from Refs.~\cite{Jensen2001_NPA695-3,
Burns2005_JPG31-S1827, Asgar2017_PRC95-031304R, Walker2009_PRC79-044321}.
The alignments $i= \langle J_x \rangle -\omega_x J_0 -\omega_x^3 J_1$
and the Harris parameters
$J_0 = 35~\hbar^2$MeV$^{-1}$ and $J_1 = 43~\hbar^4$MeV$^{-3}$
are taken from Ref.~\cite{Asgar2017_PRC95-031304R}.
}
\end{figure*}

Figure~\ref{pnc-fig2:moi} displays the comparison between experimental and calculated MOIs and
alignments for the GSBs $\pi 1/2^+[411]$ in $^{165,167,169,171}$Tm.
One can see that in general the variation of the experimental MOIs,
alignments and signature splittings with rotational frequency are
reproduced reasonably well in the PNC-CSM calculations.
In experimental data, sharp backbendings exist in the $^{165,169}$Tm nuclei
but the upbending is quite smooth and moderate in $^{167}$Tm.
Smooth upbending in $^{167}$Tm is rather well reproduced by the PNC-CSM
calculations [Figs.~\ref{pnc-fig2:moi}(b) and (f)].
 In $^{165}$Tm, the calculations predict a sharp upbending
(consistent with the backbending in experiment), and the frequency of which is close to that
of experimental backbending [Figs.~\ref{pnc-fig2:moi}(a) and (e)].
However, the PNC-CSM calculations predict a smooth and moderate upbending
instead of a sharp backbending in $^{169}$Tm [Figs.~\ref{pnc-fig2:moi}(c) and (g)].
Note that in the calculations the alignment process is more smooth in $^{171}$Tm as compared with
$^{167}$Tm [compare Figs.~\ref{pnc-fig2:moi}(d) and (h) with Figs.~\ref{pnc-fig2:moi}(b) and (f)].
However, there are no enough experimental data to confirm these predictions.
These results are quite similar to those obtained in the TRS calculations
of Ref.~\cite{Asgar2017_PRC95-031304R}.
It should be noted that in Ref.~\cite{Asgar2017_PRC95-031304R},
the calculated interaction strength at the band crossing in $^{169}$Tm ($V_{\rm int}=10$~keV)
is smaller than that in $^{165}$Tm ($V_{\rm int}=20$~keV).
This indicates that the backbending in $^{169}$Tm is sharper than the one in $^{165}$Tm,
which is inconsistent with experimental data.

In the CSM approach, the  band crossing features
depend on the interaction strength $V_{\rm int}$ between
the configurations corresponding to 1-qp band before band crossing and
3-qp configuration after band crossing.
A sharp backbending will appear for small $V_{\rm int}$ values.
A large shell gap will also make the band crossing more smooth.
In Ref.~\cite{Asgar2017_PRC95-031304R}, a smaller interaction strength and
a smaller shell gap in $^{171}$Tm than in $^{167}$Tm are predicted by TRS calculations.
As a result, TRS calculations predict sharper upbending in $^{171}$Tm than in $^{167}$Tm.

Considering the similarity of equilibrium deformations of these nuclei
(see Table~\ref{tab:defEr}) the differences in their alignment features have to be related
to the evolution of underlying neutron single-particle structure and the changes
in the position of neutron Fermi level with the increase of neutron number.
These factors and their impact on rotational properties and
band crossing features are discussed in detail below.

\begin{figure*}[!]
\includegraphics[width=0.95\textwidth]{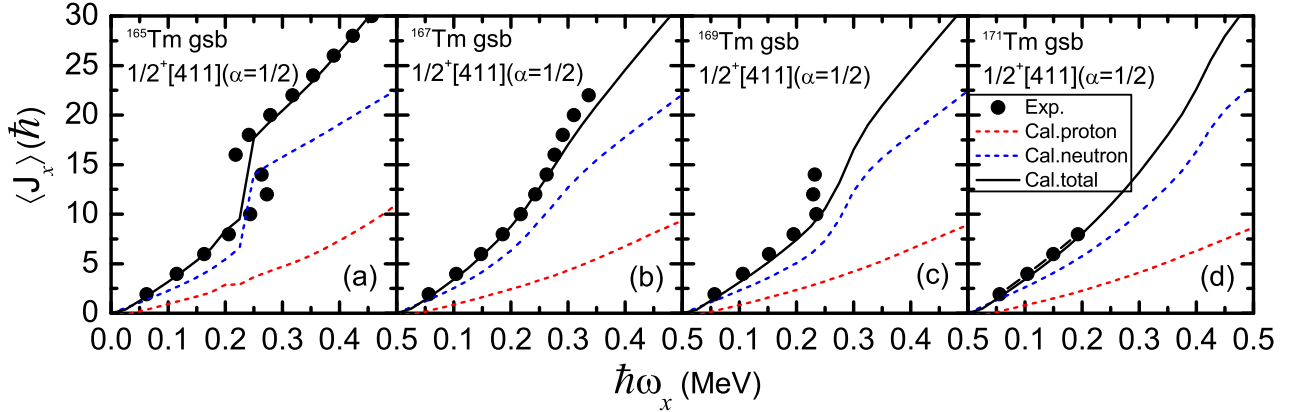}
\caption{\label{pnc-fig3:jx}
The experimental (solid circles) and calculated (black solid line) angular momentum
alignments $\langle J_x\rangle$ for the GSBs $\pi 1/2^+[411](\alpha=1/2)$
in $^{165,167,169,171}$Tm.
Proton and neutron contributions to $\langle J_x\rangle$
are shown by red and blue dashed lines, respectively.
Theoretical values are obtained in the PNC-CSM calculations.
Note that contrary to bottom panels of Fig.~\ref{pnc-fig2:moi}, smoothly increasing part
of the alignment represented by the Harris formula is not subtracted in this figure.}
\end{figure*}

Experimental and calculated angular momentum alignments $\langle J_x\rangle$
for the ground state $\pi 1/2^+ [411](\alpha=1/2)$ bands in $^{165,167,169,171}$Tm
as well as respective calculated proton and neutron contributions to
$\langle J_x\rangle$ are shown in Fig.~\ref{pnc-fig3:jx}.
Note that contrary to bottom panels of Fig.~\ref{pnc-fig2:moi}, smoothly increasing part
of the alignment represented by the Harris formula is not subtracted in Fig.~\ref{pnc-fig3:jx}.
The latter figure clearly shows that similar to even-even nuclei the first
backbendings or upbendings emerge from the band crossings in neutron subsystem.
As a result, we focus only on neutron subsystem in the discussion below.

\begin{figure*}[!]
\includegraphics[width=0.95\textwidth]{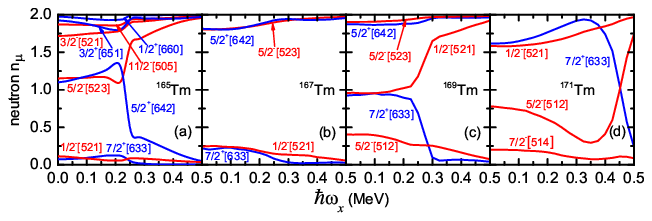}
\caption{\label{pnc-fig4:occup}
The occupation probabilities $n_{\mu}$ of neutron orbitals
$\mu$ (counting both $\alpha = \pm 1/2$ signatures together so that the
maximum occupation probability is 2.0) located close to the
Fermi level in the GSBs of $^{165,167,169,171}$Tm.
The positive (negative) parity levels are shown by blue (red) lines.
The Nilsson levels located far above (with $n_{\mu}\sim 0$) and
far below (with $n_{\mu}\sim 2.0$) the Fermi level are not shown.
}
\end{figure*}

Figure~\ref{pnc-fig4:occup} shows the occupation probabilities $n_{\mu}$
of neutron orbitals $\mu$ located close to the Fermi level in the
GSBs of $^{165,167,169,171}$Tm.
In the PNC-CSM calculations, the particle-number is
conserved from beginning to the end, whereas the occupation probabilities
$n_{\mu}$ of the orbitals change with rotational frequency.
By examining the variations of the occupation probabilities with rotational frequency,
one can get detailed insight into the backbending or upbending mechanisms.
One can see from Fig.~\ref{pnc-fig4:occup}(a) that in $^{165}$Tm at rotational frequency
$\hbar\omega_x \sim 0.25$~MeV, the occupation probability $n_{\mu}$ of the
neutron $\nu 5/2^+[642]$ orbital drops sharply from 1.4 down to 0.3,
while that of the $\nu 5/2^-[523]$ orbital increases sharply from 1.0 up to 1.7.
On the contrary, the occupation probabilities of other orbitals
(such as $\nu 3/2^-[521]$, $\nu 3/2^+[651]$ and $\nu 7/2^+[633]$) change only
modestly in the frequency range corresponding to the backbending.
This indicates that the main contribution to sharp backbending observed
in $^{165}$Tm
comes from the neutron $\nu 5/2^+[642]$ orbital emerging from the
spherical $i_{13/2}$ subshell (see also the discussion below).
 Other deformed orbitals emerging from this subshell
(such as $\nu 3/2^+[651]$ and $\nu 7/2^+[633]$) provide significantly
smaller contribution to this backbending.

In the case of $^{167}$Tm, the orbitals above (below) the Fermi level are nearly
empty (occupied) [see Fig.~\ref{pnc-fig4:occup}(b)] due to the presence of a large
shell gap at $N=98$ [see Fig.~\ref{pnc-fig1:nil}(b)]. The occupation probabilities
of the displayed orbitals are nearly constant before and after rotational frequency
range of $\hbar\omega_x=0.2-0.3$~MeV corresponding to smooth upbending in this nucleus.
The absence of sharp change of the occupation of the orbitals means that no sharp
backbending exists in $^{167}$Tm.
Gradual deoccupation of the $\nu 7/2^+[633]$ orbital and gradual occupation of
the $\nu 5/2^+[642]$ and $\nu 5/2^-[523]$ orbitals in above mentioned frequency
range is mostly responsible for the smooth upbending in this nucleus.

The situation changes in $^{169}$Tm; the occupation probability $n_{\mu}$ of
the $\nu 7/2^+[633]$ orbital decreases from 0.7 down to 0.1 and the one of the
$\nu 1/2^-[521]$ orbital increases from 1.2 up to 1.7 on going from
$\hbar\omega_x \approx 0.25$~MeV up to $\hbar\omega_x \approx 0.3$~MeV
[see Fig.~\ref{pnc-fig4:occup}(c)].
Therefore, the backbending in $^{169}$Tm comes from rapid deoccupation of the
$\nu 7/2^+[633]$ orbital.
Note that the change of the occupation probabilities $n_{\mu}$ of the
orbitals of interest in $^{169}$Tm is not as sharp as that in $^{165}$Tm and
with a higher $\Omega$ value in $\nu 7/2^+[633]$ as compared with $\nu 5/2^+[642]$,
it is understandable that the backbending in $^{169}$Tm is somewhat weaker
than in $^{165}$Tm.

For $^{171}$Tm the occupation probability $n_{\mu}$ of the $\nu 7/2^+[633]$
orbital decreases gradually from 1.9 down to 0.2, while that for the $\nu 5/2^-[512]$
orbital increases gradually from 0.3 up to 1.7 in the frequency range
$\hbar\omega_x = 0.35-0.50$~MeV.
Thus, the calculations predict a smooth upbending centered at $\hbar\omega_x\sim0.42$~MeV,
which takes places at higher frequency as compared
with backbendings/upbendings in lower $N$ Tm isotopes.

\begin{figure*}[!]
\includegraphics[width=0.95\textwidth]{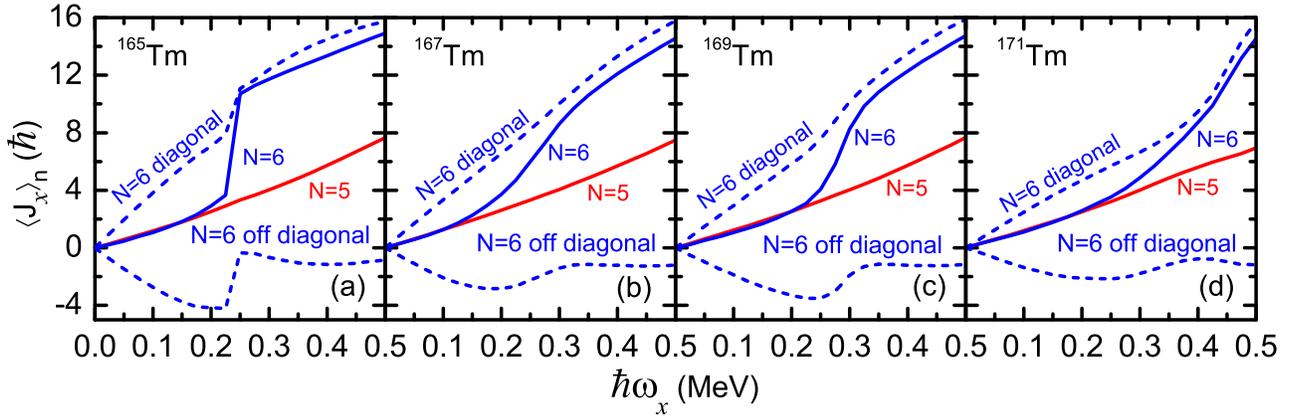}
\caption{\label{pnc-fig5:jxshell}
The contributions of neutron $N = 5, 6$ major shells to the angular
momentum alignment $\langle J_{x}\rangle$ of the GSBs
in $^{165,167,169,171}$Tm. Solid blue and red lines are used for the
$N=5$ and $N=6$ shells, respectively.
The contributions of diagonal [$\Sigma_{\mu} j_{x}(\mu)$] and off-diagonal
[$\Sigma_{\mu <\nu} j_{x}(\mu\nu)$] parts in Eq.~(\ref{eq:jx}) coming from the neutron
$N = 6$ major shell are shown by blue dashed lines.
}
\end{figure*}

The contributions of neutron $N$ = 5 and 6 major shells to the angular momentum alignment
$\langle J_{x}\rangle$ of the GSBs in $^{165,167,169,171}$Tm
are shown in Fig.~\ref{pnc-fig5:jxshell}.
In all these Tm isotopes the backbendings or upbendings emerge from the contributions
of the neutron $N=6$ major shell since at the frequencies corresponding to this phenomena
these contributions increase either drastically or gradually above the trend seen at low frequencies.
On the contrary, the $N=5$ contributions to $\langle J_{x}\rangle$ form almost straight
lines as a function of rotational frequency [see Fig.~\ref{pnc-fig5:jxshell}].
In $^{165}$Tm, sharp backbending emerges predominantly from the $N=6$ shell off-diagonal
contribution to $\langle J_{x}\rangle$;
however, smaller diagonal contribution is still present [see Fig.~\ref{pnc-fig5:jxshell}(a)].
In the case of $^{167}$Tm, smooth upbending almost fully comes from the $N=6$ shell
off-diagonal contribution to $\langle J_{x}\rangle$ [see Fig.~\ref{pnc-fig5:jxshell}(b)].
Upbending in $^{169}$Tm again dominates by the $N=6$ shell off-diagonal contribution
to $\langle J_{x}\rangle$ but relatively small diagonal contribution is
still visible [see Fig.~\ref{pnc-fig5:jxshell}(c)].
The balance of diagonal and off-diagonal contributions to $\langle J_{x}\rangle$
becomes more equal in smooth upbending of $^{171}$Tm [see Fig.~\ref{pnc-fig5:jxshell}(d)].

\begin{figure*}[!]
\includegraphics[width=0.95\textwidth]{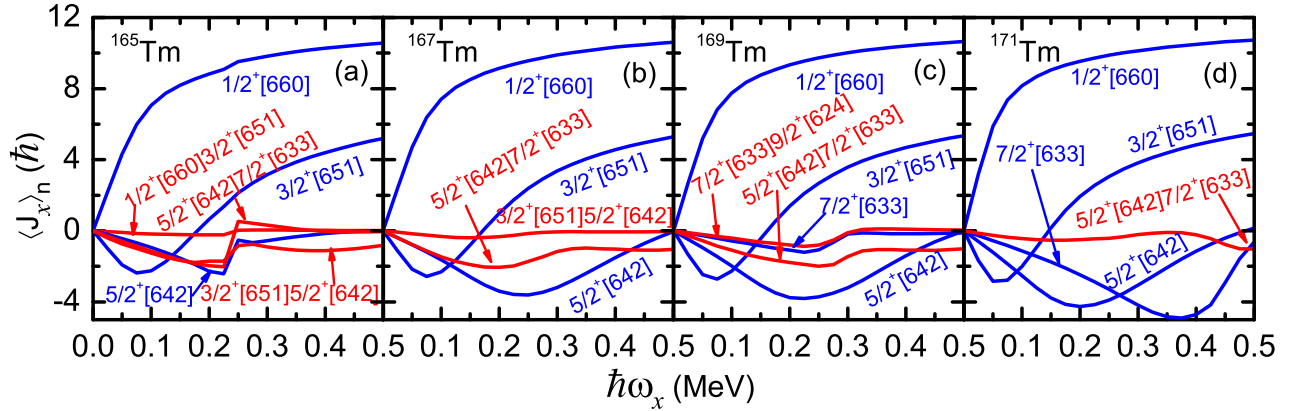}
\caption{\label{pnc-fig6:jxorb}
The contributions of the neutron $N = 6$ orbitals to the angular momentum alignments
of the GSBs in $^{165,167,169,171}$Tm.
Blue and red lines display the diagonal [$j_{x}(\mu)$] and off-diagonal [$j_{x}(\mu\nu)$]
terms in Eq.~(\ref{eq:jx}), respectively.}
\end{figure*}

In order to have a more detailed understanding of the level crossing mechanism,
the contributions of the neutron $N = 6$ orbitals to the angular momentum
alignments of the GSBs in $^{165,167,169,171}$Tm are shown
in Fig.~\ref{pnc-fig6:jxorb}.
One can see from Fig.~\ref{pnc-fig6:jxorb}(a) that
off-diagonal terms $j_x(\nu5/2^+[642]\nu7/2^+[633])$ and $j_x(\nu3/2^+[651]\nu5/2^+[642])$
and the diagonal term $ j_x(\nu5/2^+[642])$ increase drastically in the frequency region
corresponding to backbending.
This indicates that the sharp backbending in $^{165}$Tm mainly comes from these three terms.
Fig.~\ref{pnc-fig5:jxshell}(b) indicates that upbending seen at $\hbar\omega_x =0.25-0.35$ MeV
in $^{167}$Tm emerges from only $N=6$ off-diagonal terms. In the calculations,  this smooth
upbending comes only from off-diagonal term $j_x(\nu5/2^+[642]\nu7/2^+[633])$
which increases gradually in the frequency range of interest [see Fig.~\ref{pnc-fig6:jxorb}(b)].
Fig.~\ref{pnc-fig6:jxorb}(c) shows that off-diagonal terms
$j_x(\nu5/2^+[642]\nu7/2^+[633])$ and $j_x(\nu7/2^+[633]\nu9/2^+[624])$
and diagonal term $j_x(\nu7/2^+[633])$ contribute to gradual upbending in $^{167}$Tm.
One can see in Fig.~\ref{pnc-fig6:jxorb}(d) that smooth upbending in $^{171}$Tm
comes mainly from the contribution of the diagonal term $j_x(\nu7/2^+[633])$.
However, off-diagonal term $j_x(\nu5/2^+[642]\nu7/2^+[633])$
has some cancellation effects and makes the upbending in $^{171}$Tm less distinct.

Therefore, one can conclude that with increasing neutron number the Fermi level
of the Tm isotopes moves from the bottom to the top of the neutron $i_{13/2}$ subshell
and different deformed orbitals emerging from this spherical subshell contribute to the
backbendings and upbendings in these nuclei.
The backbending/upbending depends not only on the shell structure in the vicinity
of the Fermi level, but also on specific high-$j$ orbital.
With similar shell structure, higher $\Omega$ high-$j$ orbital is expected to provide
a weaker backbending/upbending as compared with small $\Omega$ high-$j$ orbital.

\begin{figure}[!]
\includegraphics[width=1.0\columnwidth]{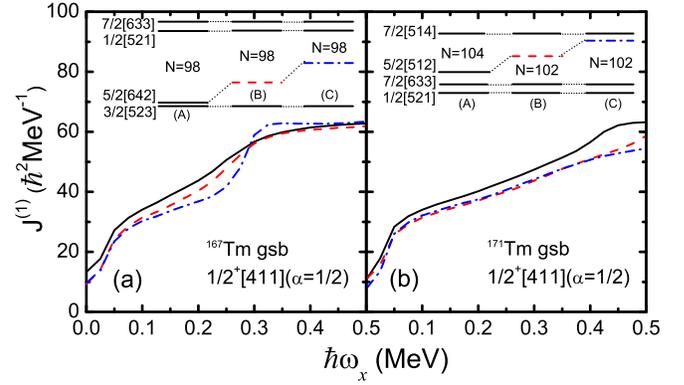}
\caption{\label{pnc-fig7:shell}
The dependence of the calculated MOIs of the GSBs
$\pi 1/2^+[411](\alpha=1/2)$ in $^{167}$Tm and $^{171}$Tm on
the size of neutron shell gaps at $N=98$ and $N=102$. Upper parts of
the panels (a) and (b) show the single-particle states at no rotation
around these gaps and their modifications discussed in the text. The
columns (A) in both panels show these states as obtained  in the
calculations with no modifications [see Fig.\ \ref{pnc-fig1:nil}(b)].
The columns (B) and (C) in panel (a) show the cases when the
neutron orbital $\nu5/2^+[642]$ is shifted up in energy by 0.04 and
0.08$\hbar\omega_0$, respectively. In the panel (b), the situations
corresponding to the shift up in energy of  the neutron  $\nu5/2^{-}$[512] orbital by
0.03 and 0.06$\hbar\omega_0$ are illustrated in the columns (B) and
(C), respectively.}
\end{figure}

In Fig.~\ref{pnc-fig7:shell} the dependence of the MOIs of selected
bands on the size of neutron gaps at $N=98$ and $N=102$ is shown with the
goal to evaluate the effects of shell gap sizes on band crossing features.
In Fig.~\ref{pnc-fig7:shell}(a), the neutron orbital
$\nu5/2^+[642]$ in $^{167}$Tm is shifted up in energy by 0.04 and 0.08$\hbar\omega_0$
to make the $N=98$  gap smaller [see Fig.~\ref{pnc-fig1:nil}(b)].
Note that $\hbar\omega_0$ is the harmonic oscillator frequency in Eq.~(\ref{eq:axialNil}).
For the latter value, the upbend in $^{167}$Tm is significantly sharper as
compared with the cases obtained for the 0.04$\hbar\omega_0$ shift
of the $\nu5/2^+[642]$ orbital and the original size of the $N=98$ gap.
In Fig.~\ref{pnc-fig7:shell}(b), the neutron  $\nu5/2^{-}$[512] orbital in $^{171}$Tm
is shifted up in energy by 0.03 and 0.06$\hbar\omega_0$ to make the $N=102$ gap larger.
It can be seen that with the $N=102$ gap increasing, smooth
upbending in $^{171}$Tm gets washed out. There is no $N=102$ gap
in our calculations without above mentioned modifications. It was suggested in
Ref.~\cite{Asgar2017_PRC95-031304R} that this may lead to a sharp backbending.
However, present calculations do not show even sharp upbend.
Thus, one can conclude that the $N=102$ gap has a smaller influence on the alignment
features as compared with the $N=98$ gap.

With increasing neutron number $N$  the neutron Fermi level moves from the vicinity
of the $\nu5/2^+[642]$ orbital towards the $\nu7/2^+[633]$ orbital. However, the gradual
alignment of the latter orbital is not affected by the size of the $N=102$ gap.
 Thus, the present calculations show that no matter whether the $N=102$ gap exists
or not, the alignment is much more gradual in $^{171}$Tm as compared
with $^{167}$Tm in which upbending is clearly visible.
Therefore, this has confirmed our previous conclusion that the
band crossing features not only depends on the shell structure close to the Fermi level,
but also on specific high-$j$ orbital located in the vicinity of this surface.

\begin{figure*}[!]
\includegraphics[width=0.95\textwidth]{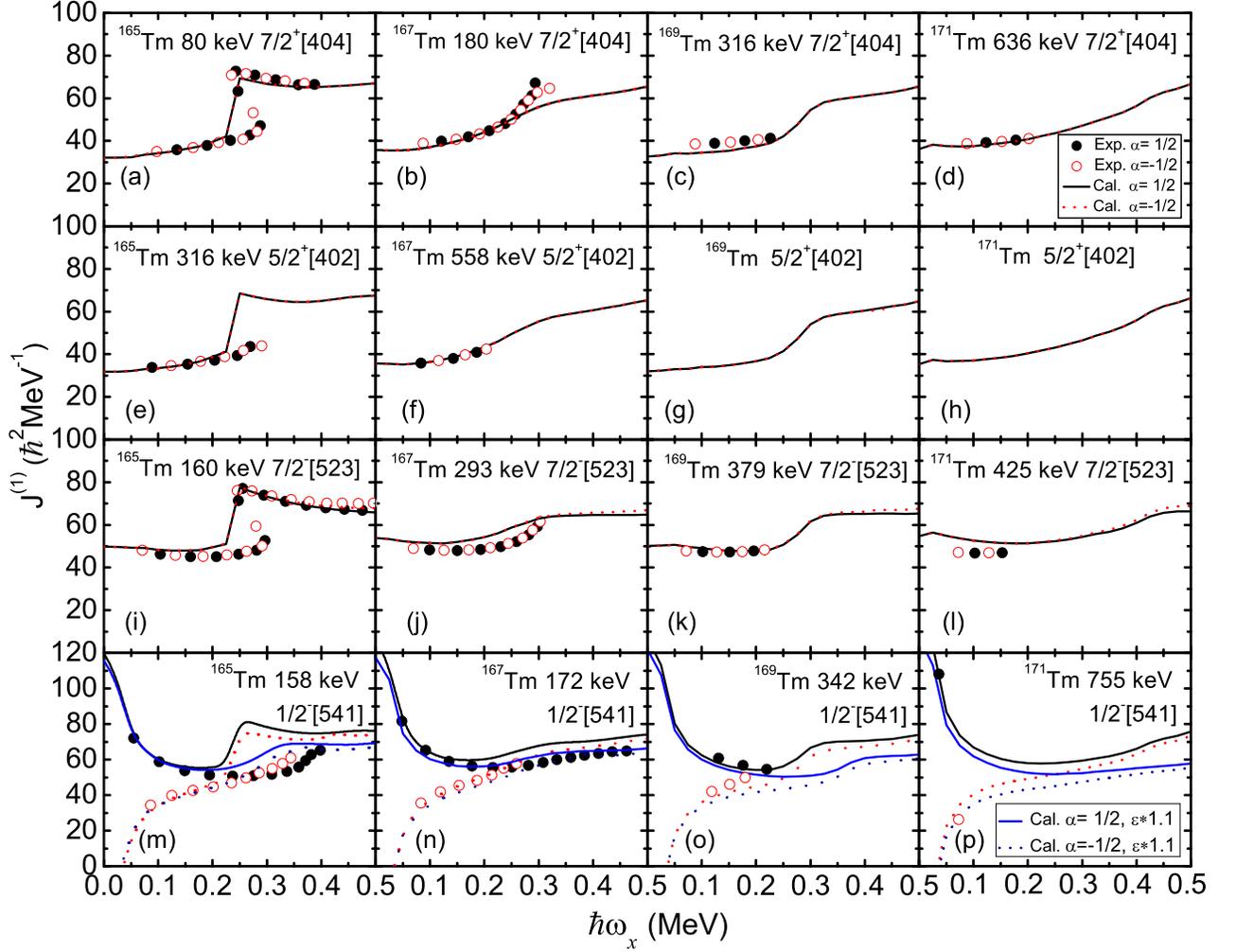}
\caption{\label{pnc-fig8:esb}
The experimental and calculated MOIs $J^{(1)}$
for excited 1-qp bands in $^{165,167,169,171}$Tm.
The experimental MOIs are displayed by
black solid and red open circles for the
$\alpha= +1/2$ and $\alpha= -1/2$ branches of rotational band, respectively.
Corresponding calculated values are shown by
black solid ($\alpha= +1/2$) and red dotted ($\alpha= -1/2$) lines.
The experimental data are taken from Refs.~\cite{Jain2006_NDS107-1075,
Baglin2000_NDS90-431, Baglin2008_NDS109-2033, Baglin2018_NDS151-334}.
The experimental excitation energies of the bandheads of
rotational bands are also displayed.
Solid and dotted blue lines are used in bottom panels for the results of the
calculations obtained with quadrupole deformation increased by 10\%.
}
\end{figure*}

There are significant number of 1-qp excited rotational bands observed in
$^{165,167,169,171}$Tm.
Fig.~\ref{pnc-fig8:esb} shows experimental and calculated MOIs for these bands.
With few exceptions, the PNC-CSM calculations reproduce their MOIs well.
For example, the PNC-CSM calculations somewhat overestimate the MOIs of the
$\pi 7/2^{-}$[523] bands in $^{167, 171}$Tm [see Figs.~\ref{pnc-fig8:esb}(j) and (l)].
In addition, the calculations predict a sharp upbending instead of backbending
seen in experiment in the $\pi 7/2^+[404]$ and $\pi 7/2^-[523]$ bands of $^{165}$Tm
[see Figs.~\ref{pnc-fig8:esb}(a) and (i)]. Similar upbending is predicted also in the
$\pi 5/2^+[402]$ band of $^{165}$Tm but it is not seen in experiment [see
Fig.~\ref{pnc-fig8:esb}(e)].

In a given nucleus, neutron configurations of the 1-qp bands are the same
because the equilibrium  deformations are the same for all bands in the
calculations. As a consequence, neutron MOIs and calculated neutron
backbending/upbending are the same for all bands; the minor
differences between calculated curves seen in panels (a,e,i), (b,f,j), (c,g,k)
and (d,h,l)  of Fig.~\ref{pnc-fig8:esb}  are due to odd proton state.
The systematics of  experimental data in this mass region shows that with  exception
of the $\pi 1/2^{-}$[541] band the backbending/upbendings frequencies for all 1-qp
rotational bands in a given nucleus are very close to each other.
In the $\pi 1/2^{-}$[541] bands the
upbending takes place in experiment at higher frequency as
compared with other bands [see Fig.~\ref{pnc-fig8:esb}(m)]  or is even
absent [see Fig.~\ref{pnc-fig8:esb}(n)].

Delayed crossing frequency in the $\pi 1/2^{-}$[541] band is explained
by strong prolate deformation driving effect of underlying single-particle
orbital; this effect has been confirmed both in experiment and in the
calculations~\cite{Nazarewicz1990_NPA512-61, Warburton1995_NPA591-323}.
Indeed, the PNC-CSM calculations with the deformation which is the same as for other
bands fail to reproduce experimental band crossing features [see
Figs.~\ref{pnc-fig8:esb}(m) and (n)]. However, the increase  of the quadrupole
deformations $\varepsilon_2$ of all $\pi 1/2^{-}[541]$ bands by 10\% leads
to a substantial improvement of the description of experimental situation
[see Figs.~\ref{pnc-fig8:esb}(m), (n) and (o)].
With this modification, the MOIs of the $\pi 1/2^{-}$[541] bands in $^{167,169}$Tm
can be reproduced rather well.
However, the experimental frequency of upbending in the $\alpha=1/2$ branch
of the $1/2^-[541]$ band in $^{165}$Tm is still underestimated in the
calculations.

\begin{figure}[!]
\includegraphics[width=1.0\columnwidth]{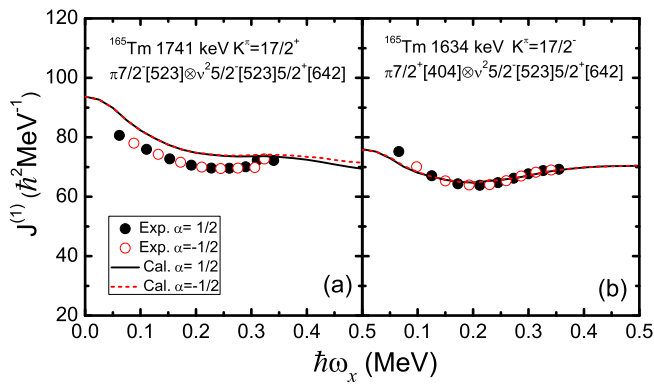}
\caption{\label{pnc-fig9:3qp}
The experimental and calculated MOIs $J^{(1)}$ of two 3-qp
bands in $^{165}$Tm with $K^\pi=17/2^+$ (a) and $K^\pi=17/2^-$ (b).
The experimental data are taken from Ref.~\cite{Jensen2001_NPA695-3}.
}
\end{figure}

Figure~\ref{pnc-fig9:3qp} shows experimental and calculated MOIs
of two 3-qp bands observed in $^{165}$Tm with the structure
$K^\pi=17/2^+$ ($\pi7/2^-[523] \otimes \nu^2 5/2^+[642]  5/2^-[523]$) and
$K^\pi=17/2^-$ ($\pi7/2^+[404] \otimes \nu^2 5/2^+[642]  5/2^-[523]$).
The MOIs of the $K^\pi=17/2^-$ band are reproduced rather well.
On the contrary, the calculated MOIs for the $K^\pi=17/2^+$ band
are larger than experimental  data due to the overestimation
of MOIs in the $\pi7/2^{-}$[523] band of this nucleus. The latter feature
is present in all Tm isotopes under study with exception of $^{169}$Tm
and it may be caused by the configuration  dependent deformation effects.

\section{Conclusions~\label{Sec:summary}}

A comparative study of three theoretical approaches, namely,
the CRHB+LN approach, the cranking CDFT-SLAP and PNC-CSM,
for the description of rotational properties has been performed using
the set of even-even and odd-$Z$ rare-earth nuclei as a testing ground.
These three models reproduce experimental MOIs (including the evolution
of MOIs with rotational frequency and band crossing features) reasonably well
for most of the cases but their predictions at rotational frequencies above
the first band crossing can differ substantially.

The comparison of these models in the case of even-even nuclei
reveals the following features:
\begin{enumerate}[(i)]
\item
There are no free parameters in the particle-hole channel of the CDFT-based models.
The calculated results obtained with different CEDFs within the framework
of one model are in general close to each other.
The results of the CRHB+LN and the cranking CDFT-SLAP calculations are
typically closer to each other than to those obtained with PNC-CSM.
Note that CDFT+LN is based on fully variational approach,
while cranking CDFT-SLAP employs shell model approach.

\item
At present, the strength of pairing correlations is adjusted to experimental
data at low spin in the CDFT-based models.
The need for that is dictated by the lack of global studies of pairing in the CDFT.
For example, the requirement for the variation of the strength of the Gogny pairing with
particle number in the RHB, CRHB and CRHB+LN approaches is known for some
time~\cite{Afanasjev2003_PRC67-024309, Wang2013_PRC87-054331, Agbemava2014_PRC89-054320}.
However, the precise form of this variation has not been established till now.
The work in that direction is in progress and there is a hope that its better definition will
allow to perform parameter free calculations of rotational properties across
the nuclear chart in the future.
Similar situation exists also in the cranking CDFT-SLAP in which the monopole pairing is used.
The implementation of separable pairing of Ref.~\cite{Tian2009_PLB676-44} into this framework
and subsequent study of particle number dependencies of separable pairing across the
nuclear chart will allow to improve the predictive power of the model.
Note that the pairing gaps obtained with the Gogny and separable pairings are closely
correlated and show the same particle number dependencies~\cite{Agbemava2014_PRC89-054320}.

\item
The models differ in the treatment of particle number.
In the cranking CDFT-SLAP, the particle number is totally conserved
from beginning to the end and the Pauli blocking effects are taken into account exactly.
By iterating the occupation probabilities of single-particle levels back to
the densities and currents, cranking CDFT-SLAP is fully self-consistent.
Due to exact particle number conservation, there is no pairing collapse even at very high spins.
On the contrary, an approximate particle number projection by means of the LN method
is employed in the CRHB+LN approach.
However, the comparison of the results of the calculations within the
cranking CDFT-SLAP and CRHB+LN approaches indicates that the LN method is a reasonable
and practical approximation to exact particle number projection.
It allows to avoid the pairing collapse for most of the cases in the frequency range
of interest making the CRHB+LN approach suitable for the description of yrast bands in
even-even systems.

\item
 However, there are some technical limitations in both CDFT-based models.
In the CRHB+LN approach, there is no numerical convergence in some cases
in the vicinity of second band crossings and at extremely high rotational frequencies.
The former numerical instability is most likely caused by the competition of
different configurations located at comparable energies in the region of
second band crossing.
The latter one is most likely connected with the pairing collapse
but it typically takes place when the additional binding due to pairing is in the
range of several tens of keV.
At these high rotational frequencies the calculations without pairing in the CRMF
framework (see Ref.~\cite{Vretenar2005_PR409-101} and references quoted therein)
represent a feasible alternative to the CRHB+LN ones
and such calculations are significantly more numerically stable.
In the cranking CDFT-SLAP calculations, minor staggering of MOIs and pairing
energies as a function of rotational frequency is observed in some
frequency ranges.
This appears for the situations of high level density of the single-particle states in the
vicinity of the Fermi level leading to competing minima corresponding
to slightly different MPCs with relatively small energy differences.
To avoid this staggering in calculated quantities the MPCs have to be
traced as a function of rotational frequency by using single-particle
level tracking technique.
However, this is extremely time consuming for systematic investigations of heavy nuclei.

\item
The PNC-CSM is built on phenomenological Nilsson potential and
employs the same treatment of pairing correlations as in cranking CDFT-SLAP.
This model is non-selfconsistent since the Nilsson
parameters ($\kappa, \mu$) are fitted to experimental data on
single-particle states and deformation parameters ($\varepsilon_2, \varepsilon_4$)
are defined from other model (such as microscopic+macroscopic model).
Therefore, the predictive power (especially in the extrapolations to other  nuclei/regions)
is lower than in the CDFT-based methods.
In addition, the PNC-CSM uses fixed deformation defined from the ground state results.
Thus, this approach faces the difficulties when the shape of nucleus
depends on the configuration (as in the case of strongly deformation
driving orbital $\pi1/2^-[541]$ in odd-mass nuclei) or when the
deformation changes with angular momentum~\cite{Simpson1984_PRL53-648, Simons2003_PRL91-162501}.

\item
 However, there are some important practical benefits of the PNC-CSM.
Its computational cost is significantly smaller as compared with
CDFT-based models and numerical calculations of odd-$A$ and even-even
nuclei require similar computational efforts (contrary to the CDFT-based
approaches which require significantly larger computational efforts in
odd-mass nuclei as compared with even-even ones).
With the Nilsson parameters ($\kappa, \mu$) fitted to experimental data on single-particle
energies, it has substantially better spectroscopic properties as compared
with CEDFs, and thus, in general, is expected to provide a better descriptive
power in odd-mass nuclei.

\item
The CDFT-based models predict sharper band crossing features as
compared with PNC-CSM calculations.
This is caused by the change of the mean fields and corresponding single-particle level structures at
the band crossing which leads to a weak interaction of the GSB and $s$-band.
In a few cases, predicted sharp upbendings contradict to experimental data.
On the contrary, the PNC-CSM predicts gradual upbendings in many cases,
which is a consequence of fixed deformation used in the calculations,
but there again are the cases contradicting to experiment.
A possible way to improve the description of band crossing features
would be to use the angular momentum projection technique instead of the cranking method.
However, as illustrated by numerous examples obtained in projected shell model that
does not necessary resolve all cases of the discrepancies between theory and
experiment~\cite{Wu2017_PRC95-064314}.
Note that a fully self-consistent mean field method with angular momentum
projection, configuration mixing and exact particle-number conservation
would be extremely time-consuming.
\end{enumerate}

 As illustrated by few examples, the calculations of rotational properties in odd-mass
nuclei are very time-consuming in the cranking CDFT-SLAP and CRHB+LN approaches
and they face a number of technical difficulties.
Thus, the systematic investigation of such properties in odd-$A$
$^{165,167,169,171}$Tm nuclei has been performed only in the PNC-CSM.
In these calculations, with few
exceptions the evolutions of the MOIs of the 1-qp and 3-qp bands with rotational
frequency (including  backbending/upbending regions) as well as their changes with
neutron number are reproduced reasonably well.
By analyzing the occupation probabilities of the Nilsson orbitals located in
the vicinity of the Fermi level and the contributions of each major shell to
the angular momentum alignments, the origin and the evolution of the
backbendings/upbendings with neutron number in these nuclei are clearly understood.
The impact of the shell gap size on the band crossing features is also investigated.
In the PNC-CSM calculations, the band crossing features depend not only on the shell
structure in the vicinity of the Fermi level but also on specific high-$j$ orbital.

\section{Acknowledgement}

Helpful discussions with  Z. Shi and B. W. Xiong are gratefully acknowledged.
This work is supported by National Natural Science Foundation of
China (11875027, 11775112, 11775026, 11775099),
Fundamental Research Funds for the Central Universities (2018MS058),
the program of China Scholarships Council (No. 201850735020)
and the U.S. Department of Energy, Office of Science,
Office of Nuclear Physics under Award No. DE-SC0013037.

%

\end{document}